\documentclass[reqno,draft]{amsart}
\usepackage{multicol, color}
\newcommand{\rem}[1]{}

\theoremstyle{plain}
\newtheorem{lemma}{Lemma}
\newtheorem{theorem}[lemma]{Theorem}

\newtheorem{definition}[lemma]{Definition}
\theoremstyle{remark}
\newtheorem{remark}{Remark}

\newcommand*{\nablab}{{\bf\nabla}}
\newcommand*{\uhat}{{\hat u}}

\newcommand*{\ubar}{{\bar u}}
\newcommand*{\ubarbf}{{\bf\bar u}}
\newcommand*{\Ubar}{{\bar U}}
\newcommand*{\pbar}{{\bar p}}
\newcommand*{\Pbar}{{\bar P}}
\newcommand*{\vbar}{{\bar v}}

\newcommand*{\ang}[1]{\left\langle #1 \right\rangle}

\def\ee{\epsilon}
\def\aa{\alpha}

\def\dd{\delta}

\def\ss{\sigma}

\def\Dd{\Delta}

\def\Om{\Omega}

\def\pp{\partial}

\begin{document}
\title[Global Well-posedness of
 Simplified Bardina Turbulence Model] {Global
Well-posedness of the Three-dimensional Viscous and Inviscid
Simplified Bardina Turbulence Models}
\date{August 9, 2006}

\author[Y. Cao]{Yanping Cao}
\address[Y. Cao]
{Department of Mathematics\\
University of Ca, Irvine\\
}
\email{ycao@math.uci.edu}
\author[E. Lunasin]{Evelyn M. Lunasin}
\address[E.M. Lunasin]
{ Department of Mathematics\\
University of California \\
Irvine \\
}
\email{emanalo@math.uci.edu}
\author[E.S. Titi]{Edriss S. Titi}
\address[E.S. Titi]
{Department of Mathematics \\
and  Department of Mechanical and  Aerospace Engineering \\
University of California \\
Irvine, CA  92697-3875, USA \\
{\bf ALSO}  \\
Department of Computer Science and Applied Mathematics \\
Weizmann Institute of Science  \\
Rehovot 76100, Israel}
\email{etiti@math.uci.edu and edriss.titi@weizmann.ac.il}

\begin{abstract}
In this paper we present analytical studies of  three-dimensional
viscous and inviscid simplified Bardina turbulence models with
periodic boundary conditions.  The global existence and uniqueness
of weak solutions to the viscous model has already been established
by Layton and Lewandowski. However, we prove here the global
well-posedness of this model for weaker initial conditions. We also
establish an upper bound to the dimension of its global attractor
and identify this dimension with the number of degrees of freedom
for this model. We show that the number of degrees of freedom of the
long-time dynamics of the solution is of the order of
$(L/l_d)^{12/5}$, where $L$ is the size of the periodic box and
$l_d$ is the dissipation length scale-- believed and defined to be
the smallest length scale actively participating in the dynamics of
the flow. This upper bound estimate is smaller than those
established for Navier-Stokes-$\alpha$, Clark-$\alpha$ and
Modified-Leray-$\alpha$ turbulence models which are of the order
$(L/l_d)^{3}$.  Finally, we establish the global existence and
uniqueness of weak solutions to the inviscid model. This result has
an important application in computational fluid dynamics when the
inviscid simplified Bardina model is considered as a regularizing
model of the three-dimensional Euler equations.
\end{abstract}
\maketitle

\\


\section{Introduction}   \label{S-1}
 Let us denote by $v(x,t)= (v_1(x,t),v_2(x,t), v_3(x,t))$ the velocity field of an incompressible fluid and $p(x,t)$ its pressure.  The three-dimensional (3D) Navier-Stokes equations (NSE)
\begin{equation}\label{NSE}
\aligned
\pp_t v  -\nu \Delta v+ \nabla\cdot(v \otimes v) &= -\nabla p + f,\\
\nabla \cdot v &= 0,\\
v(x,0) &= v^{in}(x),
\endaligned
\end{equation}
governs the dynamics of homogeneous incompressible fluid flows, where $f(x) =(f_1(x),f_2(x),f_3(x))$  is the body force assumed, for simplicity, to be time independent.  The existing mathematical theory and techniques are not yet sufficient to prove the global well-posedness of the 3D NSE. Researchers who are investigating this question have incorporated the use of computers to analyze the dynamics of turbulent flows by studying the direct numerical simulation (DNS) of these equations.  However, this is still a prohibitively expensive task to perform even with the most technologically advanced state-of-the-art computing resources.  Tracking the pointwise flow values by numerical simulation for large Reynolds number is not only difficult but also, in some cases, disputable due to sensitivity of numerical solutions to perturbation errors in the data and the limitations of reliable numerical
resolution.  In many practical applications, knowing the mean characteristics of the flow by averaging techniques is sufficient.  However, averaging the nonlinear term in NSE leads to the well-known closure problem.\\\\
To be more precise, if ${\bar{v}}$ denotes the filtered/averaged velocity field then the Reynolds averaged NSE (RANS)
\begin{equation}\label{RANS}
\aligned
\pp_t \bar{v}  -\nu \Delta \bar{v}+ \nabla\cdot(\overline{v\otimes v}) &= -\nabla\bar{p} + \bar{f},\\
\nabla \cdot \bar{v} &= 0,
\endaligned
\end{equation}
where
\begin{equation}
\begin{split}
  \nabla\cdot(\overline{v\otimes v}) &= \nabla\cdot(\bar{v}\otimes \bar{v}) + \nabla\cdot \mathcal{R}(v,v),\\
  \mathcal{R}(v,v) &= \overline{v\otimes v} -\bar{v}\otimes \bar{v}
\end{split}
\end{equation}
is not closed.  The quantity $\mathcal{R}(v,v)$ is known as the
Reynolds stress tensor. The RANS system of equations contains the
unknown quantity $\tilde{v}= v - \bar{v}$, which represents the
fluctuation around the filtered velocity $\bar{v}$. The equation in
(\ref{RANS}) is not closed because we cannot write it in terms of
$\bar{v}$ alone. The main essence of turbulence modeling is to
derive simplified, reliable and computationally realizable closure
models.
\\\\
In 1980, Bardina et al. \cite{Bardina} suggested a particular closure model by approximating  the Reynolds stress tensor by
\begin{equation}\label{Bardina-tensor}
\mathcal{R}(v,v) \approx \overline{\bar{v}\otimes \bar{v}} -\bar{\bar{v}}\otimes \bar{\bar{v}}.
\end{equation}
In \cite{Layton-06}, Layton and Lewandowski considered a simpler approximation of the Reynolds stress tensor, given by
\begin{equation}
\mathcal{R}(v,v) \approx \overline{\bar{v}\otimes \bar{v}} -\bar{v}\otimes\bar{v}.
\end{equation}
This is equivalent form to the approximation
\begin{equation}
\nabla\cdot(\overline{v\otimes v}) \approx \nabla \cdot(\overline{\bar{v}\otimes \bar{v}}).
\end{equation}
Hence, Layton and Lewandowski studied the following sub-grid scale turbulence model:
\begin{equation}\label{Layton-model}
\begin{split}
w_t -\nu \Delta w+ \nabla\cdot(\overline{w\otimes w}) &=- \nabla q + \bar{f},\\
\nabla\cdot w &= 0,\\
w(x,0) &= \bar{v}_0(x),
\end{split}
\end{equation}
where they denoted $(w,q)$, the approximation to
$(\bar{v},\bar{p})$. In this paper we will call this particular
model the {\it simplified Bardina model.}  Similar to the alpha
models \cite{CH98, CH99,CH00,FHTM,CHOT,ILT}, Layton and Lewandowski
\cite{Layton-06} used the smoothing kernel associated with the
Helmholtz operator $(I-\aa^2\Delta)^{-1}$. That is, if $v$ denotes
the unfiltered velocity and $u$ denotes the smoothed filtered
velocity, then we have the relationship $v=u-\alpha^2 \Delta u$. For
abstract mathematical study, one can define a more general smoothing
kernels, which gives a different relationship between $u$ and $v$
(see, e.g., \cite{BIL}, \cite{OlTi}).  In this paper, we will keep
the same exact smoothing operator.  There is a very important reason
behind the choice of this particular smoothing kernel in our
mathematical studies.  The reason can be traced back from the early
study of 3D Navier-Stokes-$\alpha$ (NS-$\alpha$) turbulence model
(also known as the viscous Camassa-Holm equations (VCHE) and
Lagrangian averaged Navier-Stokes-$\alpha$ (LANS-$\alpha$) model ).
The explicit analytical steady state  solutions to the NS-$\alpha$
model were found to compare successfully with empirical data for
mean velocity and Reynolds stresses for turbulent flows in channels
and pipes for wide range of Reynolds numbers (see, e.g.,
\cite{CH98,CH99,CH00}). It was, in fact, this important finding,
which led the authors of \cite{CH98,CH99,CH00}  to suggest that the
3D NS-$\alpha$ model be used as closure model for the Reynolds
averaged equations (RANS). Under this particular relationship
$v=u-\alpha^2 \Delta u$ between $u$ and $v$, the other alpha models
reduced under the channel and pipe symmetry yield exactly the same
equations, up to a modified pressure, to the system of equations for
the NS-$\alpha$ model restricted to this symmetry.  Hence, the
explicit steady state solution to these equations will match the
experimental data as well. This is one important property shared by
all the alpha models.  In particular, the simplified Bardina model
enjoys this important property as well. A more detailed discussion
of this will be presented in section \ref{closure}.  With this at
hand, we can rewrite the simplified Bardina model
(\ref{Layton-model}) as
\begin{equation}\label{Bardina}
\aligned
\pp_t v  -\nu \Delta v+ (u\cdot\nabla) u &= -\nabla p + f,\\
\nabla \cdot u &=  \nabla \cdot v = 0,\\
v &= u-\aa^2\Dd u,\\
u(x,0) &= u^{in}(x)\\
u \mbox{ and } v  \mbox{ are periodic, with periodic box } \Omega &= [0,2\pi L]^3
\endaligned
\end{equation}
Notice that consistent with all the other alpha models, the above system is the Navier-Stokes system of equations when
$\aa = 0$, i.e. $u = v$.  We have rewritten equation (\ref{Layton-model}) in the particular form (\ref{Bardina})
in order to coordinate its similarity with the family of alpha models \cite{CH98,CH99,CH00,CHOT,CHTi,FHTM,HM, ILT}.
In this form, when compared to the other alpha sub-grid scale turbulence models, the main difference, namely in the bilinear term, can be distinguished easily.

Moreover, we note that, in addition to the remarkable match, in the
channels and pipes, of explicit analytical steady state solutions of
the alpha models to the experimental data  the validity of the first
alpha model, the NS-$\alpha$ model, as a subgrid scale turbulence
model was also tested numerically in \cite{CH01} and \cite{MM}. In
the numerical simulation of the 3D NS-$\alpha$ model, the authors of
\cite{CH01}, \cite{GH}, \cite{GH2} and \cite{MM} showed that the
large scale (to be more specific, those scales of motion bigger than
the length scale $\alpha$) features of a turbulent flow is captured.
Then, for scales of motion smaller than the length scale $\alpha$,
the energy spectra decays faster
 in comparison to that of NSE.
This numerical observation has been justified analytically in \cite{FHTP}.
In direct numerical simulation, the fast decay of the energy spectra for scales of motion
smaller than the supplied filter length represents reduced grid requirements in simulating a flow.
The numerical study of \cite{CH01} gives the same results.  The same results hold as well in the study of the
Leray-$\alpha$ model in \cite{CHOT} and \cite{GH}.

This paper is arranged as follows.  In section \ref{Def-1} we fix some notations and define the functional setting.    In section \ref{closure} we discuss in further details why we chose the particular smoothing kernel and justify the use of the simplified Bardina model as a closure model to the RANS. In section \ref{E-U} we will re-establish the global existence and uniqueness of weak
solutions of equation (\ref{Bardina}) subject to periodic boundary conditions.  We will
re-establish this result requiring a weaker initial condition than those required in \cite{Layton-06}.  In section \ref{S-A}
we also provide an upper bound to the dimension of its global attractor.  We then relate this upper bound to the
number of degrees of freedom of the long-time dynamics of the solutions to this model.  Our results show that the number of
degrees of freedom for this model is proportional to $(L/l_d)^{12/5}$.  This estimate is much smaller compared to
those established for the 3D Clark-$\alpha$ \cite{CHTi}, 3D NS-$\alpha$ model \cite{FHTM} and the 3D
Modified-Leray-$\alpha$ model \cite{ILT} which are of the
order $(L/l_d)^{3}$.  The smaller estimate on the number
of degrees of freedom
for the simplified Bardina model is expected since it
has a milder nonlinear term than the 3D Clark-$\alpha$
model, 3D NS-$\alpha$ model and the 3D
Modified-Leray-$\alpha$ model.  Notice, however,
that we have excluded the Leray-$\alpha$ model in our comparison above.  For the Leray-$\alpha$ model, the estimate for its number of degrees of freedom is of the order $(L/l_d)^{12/7}$ as shown in \cite{CHOT}.  The power (12/7) is smaller than the power (12/5) of our estimate on the simplified Bardina model even though we have here a smoother nonlinear term $u\cdot\nabla u$ compared to nonlinear term $u\cdot\nabla v$ of the Leray-$\alpha$. One reason for this is that the energy dissipation length
scale $l_d$ for the Leray-$\alpha$ model is different
from the $l_d$ of the simplified Bardina, 3D NS-$\alpha$, Clark-$\alpha$ and Modified Leray-$\alpha$ model.  For the Leray-$\alpha$ model, the dissipation length
scale $l_d$ is based on the time average of the $H^3-$ norm of $u$.  On the other hand, the $l_d$ of the simplified Bardina, 3D NS-$\alpha$, Clark-$\alpha$ and Modified Leray-$\alpha$ model is based on the time average of the $H^2-$ norm of $u$. Recently, Holm and Gibbon \cite{Holm-Gibbon} produced an interpretation of the dimension of the global attractor in terms of the Reynolds number.  This global interpretation can assist in making across the broad comparison between the various alpha models.  In particular, by following their work one would be able to show that the dimension of the global attractor for the simplified Bardina model is much smaller than that of the NS-$\alpha$ model, but larger than that of the Leray-$\alpha$.

For completeness, in section \ref{ES} we also include in our study the energy spectra of the simplified Bardina model.
Although the dimension of the global attractor for the simplified Bardina model is smaller in comparison to those
established for the 3D NS-$\alpha$ model and the 3D Modified-Leray-$\alpha$ model, we found that the spectral slopes
for the energy spectra for the simplified Bardina model is the same to that of 3D Clark-$\alpha$, 3D NS-$\alpha$ model and the 3D
Modified-Leray-$\alpha$ established in \cite{CHTi,FHTM,ILT} respectively.

In the last section we prove the global existence and uniqueness of
the inviscid simplified Bardina model.  This result has important
consequences in computational fluid dynamics when the inviscid
simplified Bardina model is considered as a regularizing model of
the 3D Euler equations.  This is because the inviscid simplified
Bardina is globally well-posed model that approximates the 3D Euler
equations without adding any hyperviscous regularizing terms.  In
particular, we propose the inviscid simplified Bardina model as a
tool for testing claims about the formation of a finite time
singularity in the 3D Euler equations (see, e.g., \cite{Hou},
\cite{Kerr}  and references therein).

\section{Functional Setting and Preliminaries}\label{Def-1}

Let $\Omega = [0,2\pi L]^3$.  The simplified Bardina turbulence model
(\ref{Bardina}) of viscous incompressible flows, subject to periodic boundary
condition, with basic domain $\Omega$, is written in expanded form:
\begin{equation}\label{DLE}
\aligned
\pp_t(u-\aa^2\Dd u) - \nu\Dd (u-\aa^2\Dd u) &+ (u\cdot\nabla)
 u = -\nabla p + f,\\
 \nabla \cdot u& = 0,\\
u(x,0)& = u^{in}(x),
\endaligned
\end{equation}
where, $u$ represents the unknown ``filtered'' fluid velocity
vector, and $p$ is the unknown ``filtered'' pressure scalar; $\nu >
0$ is the constant kinematic viscosity, $\aa > 0$ is a length scale
parameter which represents the width of the filter.  The function
$f$ is a given body forcing assumed, for the simplicity of our
presentation, to be time independent and with mean zero, that is
$\int_\Omega f(x) dx = 0$, and $u^{in}$ is the given initial
velocity also assumed to have zero mean and hence the solutions $u$
and $v$ as well.

Next, we introduce some preliminary background material following the usual
notation used in the context of the mathematical theory of Navier-Stokes
equations (NSE) (see, e.g., \cite{CF88,TT84, TT88}).
\begin{enumerate}
\item We denote by $L^p$ and $H^m$ the usual Lebesgue and Sobolev spaces,
respectively.  And we denote by $|\cdot|$ and $(\cdot,\cdot)$ the $L^2-$norm
and $L^2-$inner product, respectively.
\item Let $\mathcal{F}$ be the set of all vector trigonometric polynomials
with periodic domain $\Omega$.  We then set
$$
\mathcal{V}=\left\{\phi \in
\mathcal{F}:\nabla\cdot\phi = 0 \ \mbox{and}
\int_\Omega \phi(x)\ dx = 0\right\}.
$$
We set $H$ and $V$ to be the closures of $\mathcal{V}$ in $L^2$ and $H^1$,
respectively.  We also note that by Rellich lemma (see, e.g., \cite{AR75}) we have the $V$ is compactly embedded in $H$.
\item We denote by $P_\ss:L^2 \rightarrow H$ the Helmholtz-Leray orthogonal
projection operator, and by $A = -P_\ss\Dd$ the Stokes operator subject to
periodic boundary condition with domain $D(A) = (H^2(\Omega))^3\cap V$.  We
 note that in the space-periodic case,
\begin{equation*}
Au = -P_\ss\Dd u = -\Dd u, \hspace{.5cm} \mbox{for all }u \in D(A).
\end{equation*}
The operator $A^{-1}$ is a self-adjoint positive definite compact operator
from $H$ into $H$. (cf. \cite{CF88, TT84}).  We denote by
$0 < {L}^{-2}= \lambda_1 \leq \lambda_2 \leq \dots \dots$ the
eigenvalues of $A$, repeated according to their multiplicities.  It is well
known that in three dimensions , the eigenvalues of the operator $A$ satisfy
the Weyl's type formula (see, e.g., \cite{Agmon,CF88,MTV,TT88}) namely, there
exists a dimensionless constant $ c_0 > 0$ such that
\begin{equation}\label{asymptotic}
\dfrac{j^{2/3}}{c_0} \leq \dfrac{\lambda_j}{\lambda_1} \leq c_0 j^{2/3}, \hspace{.5 cm}\mbox{for } j = 1,2,\dots,.
\end{equation}
We also observe that,  $D(A^{n/2}) = (H^n(\Omega))^3\cap V$.
\item We recall the following three-dimensional interpolation and Sobolev
inequalities (see, e.g., \cite{AR75} and \cite{CF88}):
\begin{equation}\label{sobolev}
\aligned
&\|\phi\|_{L^3}\leq c\|\phi\|_{L^2}^{1/2}\|\phi\|_{H^1}^{1/2},
\hspace{.5cm} \mbox{and }\\
&\|\phi\|_{L^6}\leq c\|\phi\|_{H^1}, \hspace{.5cm} \mbox{for every }
 \phi \in H^1(\Omega).
\endaligned
\end{equation}
Also, recall the Agmon's inequality (see, e.g., \cite{Agmon,CF88}):
\begin{equation}\label{Agmon}
\|\phi\|_{L^\infty}\leq c\|\phi\|^{1/2}_{H^1}\|\phi\|_{H^2}^{1/2},
\hspace{.5cm}\mbox{for every } \phi \in H^2(\Omega).
\end{equation}
Hereafter $c$ will denote a generic dimensionless constant.
\item For $w_1, w_2 \in \mathcal{V}$, we define the bilinear form
\begin{equation}\label{B1}
B(w_1,w_2) = P_\ss((w_1\cdot\nabla)w_2).
\end{equation}
\end{enumerate}
In the following lemma, we will list certain relevant inequalities and properties of $B$ (see \cite{CF88,TT84}).
\begin{lemma}\label{lem1-buvw}
The bilinear form B defined in (\ref{B1}) satisfies the following:
\begin{enumerate}
\item $B$ can be extended as a continuous map $B:V \times V \rightarrow V'$,
where $V'$ is the dual space of $V$.  In particular, for every $w_1,w_2,w_3 \in V$,
the bilinear form $B$ satisfies the following inequalities:
\begin{eqnarray}\label{buvw}
|\ang{B(w_1,w_2),w_3}_{V'}| \leq c|w_1|^{1/2}\|w_1\|^{1/2}\|w_2\|\|w_3\|,
\end{eqnarray}
\begin{eqnarray}\label{buvw1}
|\ang{B(w_1,w_2),w_3}_{V'}| \leq c\|w_1\|\|w_2\| |w_3|^{1/2}\|w_3\|^{1/2}.
\end{eqnarray}
Moreover, for every $w_1,w_2,w_3 \in V$, we have 
\begin{equation}\label{bilinear}
\ang{B(w_1,w_2),w_3}_{V'} = -\ang{B(w_1,w_3),w_2}_{V'}.
\end{equation}
And in particular,
\begin{equation}\label{buww}
\ang{B(w_1,w_2),w_2}_{V'} = 0.
\end{equation}
\item For $w_1\in V$ and $w_3\in D(A)$, we have
\begin{equation}\label{buvw2}
|\ang{(B(w_1,w_1),w_3)_{V'}}|= | \ang{(B(w_1,w_1),w_3)_{D(A)'}}|
\leq \lambda_1^{-1/4}|Aw_3||w_1|\|w_1\|,
\end{equation}
where $D(A)'$ is the dual space of $D(A)$.
\end{enumerate}

\end{lemma}

Using the bilinear form $B$ and the linear operator $A$, the sytems in (\ref{Bardina}) and (\ref{DLE}) is equivalent to the functional differential equation
\begin{equation}\label{Bardina-func}
\begin{split}
\dfrac{dv}{dt} + \nu Av + B(u,u) &= f,\\
v &= u + \aa^2Au,\\
v(0) = v^{in} &= u^{in} + \aa^2 Au^{in},
\end{split}
\end{equation}
\begin{definition}
{\bf(Weak Solution)}  Let $f \in H$, $u(0)=u^{in}$ $\in$ V, and $T > 0$.  A function $u \in C([0,T];V)\cap L^2([0,T];D(A))$ with $\dfrac{du}{dt}\in L^2([0,T];H)$
is said to be a weak solution to (\ref{Bardina-func}) in the interval $[0,T]$ if it satisfies the following:
\begin{equation}\label{W-soln}
\ang{\dfrac{dv}{dt},w}_{D(A)'} + \nu\ang{Av,w}_{D(A)'}+ \left(B(u,u),w\right) = (f,w),
\end{equation}
for every $w\in D(A)$.
Here, the equation (\ref{W-soln}) is understood in the following sense:
\\
For almost everywhere $t_0,t\in [0,T]$ we have
\begin{equation}
\ang{v(t), w}_{V'}-\ang{v(t_0),w}_{V'} + \nu\int_{t_0}^t(v,Aw)+\int_{t_0}^t \left(B(u(s),u(s)),w\right) ds = \int_{t_0}^t(f,w)ds.
\end{equation}
\end{definition}

\section{The simplified Bardina model as a turbulence closure model}\label{closure}
As we mentioned earlier, one important characteristic shared by all the alpha models is the particular kernel used to give the relation between the smoothed velocity $u$ and unsmoothed velocity $v$.  This particular choice of smoothing kernel gives the important result that under the pipe and channel symmetry, the reduced equation of all the other alpha models takes the form of the reduced of NS-$\alpha$ under the same symmetry, up to modified pressure.  As a
result, the explicit analytical steady state solutions
to these equations will resemble
the explicit analytical steady state solutions of the NS-$\alpha$.  In this way, the excellent
match of explicit analytical steady state solutions
of NS-$\alpha$ to experimental data in the channel
and pipe symmetry for wide range of
Reynolds number (\cite{CH98,CH99,CH00}) is also
inherited by these models.  In this section, we consider
the simplified Bardina model as a closure to the
stationary Reynolds averaged Navier-Stokes (RANS) equations.  We will show that the reduction of the system of equations in (\ref{DLE}) or (\ref{Bardina}) in the infinite channels and pipes are the same (up to modified pressure) as the system of equations obtained in the case of NS-$\alpha$ (or the viscous Camassa-Holm equations (VCHE)), \cite{CH98, CH99, CH00}. \\
Let us begin by recalling the stationary RANS equations
in channels
and pipes (see, e.g., \cite{Pope,Townsend}). We establish some
notations:  for a given function $\phi(x,t)$ we denote by
\begin{equation}
\ang{\phi}(x) = \bar{\phi}(x) = \lim_{T\rightarrow\infty}\dfrac{1}{T}\int_0^T\phi(x,t)dt
\end{equation}
assuming that such a limit exists (see, e.g., \cite{FMRT} for the
generalization of the notion of limit to make sense of infinite time
averages.)
 The long (infinite) time average of the NSE, i.e. the stationary RANS
 equations, are given by
\begin{equation}\label{RANS}
\aligned
(\ubarbf\cdot\nablab)\ubarbf& = \nu\Dd \ubarbf -
 \nablab \pbar-\overline{({\bf u}-\ubarbf)\cdot\nablab({\bf u}-\ubarbf)}\\
\nablab\cdot\ubarbf &= 0.
\endaligned
\end{equation}
This averaging process yields the well known closure problem. The
system above is not closed since we cannot express it solely in
terms of $ \ubarbf$ alone.  The main   idea behind turbulence
modeling is to produce an approximate closed form for (\ref{RANS})
in terms of $\ubarbf$ alone.

\subsection{The RANS equations  for Turbulent Channel Flows}
As might be expected from the visual appearance of the flow in experimental observations of turbulent Poiseuille flows in infinite channel (see, e.g., \cite{Pope,Townsend}), the mean velocity in (\ref{RANS}) for turbulent channel flows takes the form ${\bf\ubar} = \left[\Ubar(z),0,0\right]^T$, where $\Ubar(z) = \Ubar(-z)$, with mean pressure $\pbar= \Pbar(x,y,z)$. Using this classical observation, the RANS system \eqref{RANS} under such symmetry reduces to:
\begin{equation}\label{RANSCHANNEL}
\aligned
-\nu \Ubar''+\pp_z\ang{wu} &= -\pp_x \Pbar\\
\pp_z\ang{wv} &=  -\pp_y\Pbar\\
\pp_z\ang{w^2} &=-\pp_z\Pbar
\endaligned
\end{equation}
where the prime ($'$) denotes the derivative in the $z$-direction,
and $(u,v,w)^T = {\bf u-\ubar }$ is the fluctuation of the velocity
in the infinite channel $\{(x,y,z)\in \mathbb{R},-d\leq z\leq d\}$.
It is also observed from the experiments (see, e.g. \cite{Pope,Townsend}), that the Reynolds stresses $\ang{wu}, \ang{wv} \mbox{and }\ang{w^2}$ are functions of the
variable $z$ alone.  At the boundary, it is natural to impose the conditions
$\Ubar(\pm d) = 0$ (no-slip) and $\nu \Ubar'(\pm d) = \mp \tau_0$,
where $\tau_0$ is the boundary shear stress.  Using the boundary
conditions $\ang{wu}(\pm d)=\ang{wv}(\pm d)=0$, the Reynolds equations
imply that $\ang{wv}=0$ and $\Pbar = P_0 -\tau_0x/d - \ang{w^2}(z)$,
with integration constant $P_0$.
\subsection{The Reduced Simplified Bardinal Model for Channel Flows}
For any turbulence model if an explicit analytical solution is
available, then one can match this solution with the available
physical experimental data to test its validity.   Here we will show
that the reduced simplified Bardina model under the channel symmetry
admits the same exact equation as the reduced NS-$\aa$ model.  This
is enough to show that the numerical solution of the reduced
simplified Bardina model in the channel will match the experimental
data for wide range of Reynolds number.   For the simplified Bardina
system of equations, under the channel symmetry, we denote by {\bf
U} the velocity $u$ in (\ref{DLE}) and we seek its steady state
solutions in the form ${\bf U}=\left[U(z),0,0\right]^{T}$, with even
reflection symmetry condition $U(z) = U(-z)$, and boundary condition
$U(\pm d)=0$. Under these conditions, the steady simplified Bardina
equations reduces to:
\begin{equation}\label{SBM-channel}
\aligned
-\nu V'' = -\nu U'' + \nu \aa^2U'''' &= -\pp_x p\\
0 &= -\pp_y p\\
0 &= -\pp_z p
\endaligned
\end{equation}
where $V = U-\aa^2U''$ and $p$  is a pressure function. Notice here that we need
additional boundary conditions to determine $V$.  Such boundary conditions are
not yet available based on physical considerations.  However, in this case, and
under the symmetry of the channel, the missing boundary conditions come as free
parameters that will be determined through a  tuning process with  empirical
data.
\subsection{Identifying the Simplified Bardina Model with RANS - The Channel case}\label{SBMRANS}
Following the idea of \cite{CH98,CH99,CH00} we identify the systems (\ref{RANSCHANNEL})
and (\ref{SBM-channel}) with each other, which is the essence of our closure assumption.
We compare (\ref{RANSCHANNEL}) and (\ref{SBM-channel}), and as a result, we identify the
various counterparts as
\begin{equation}\label{SBM-alphavsRANS}
\aligned
\Ubar &= U \\
\pp_z\ang{wu}&= \nu\aa^2U''''+p_1\\
\pp_z\ang{wv}&= 0\\
\nabla(\Pbar+\ang{w^2})&= \nabla(p-p_1x)\\
\endaligned
\end{equation}
for some constant $p_1$.  This identification gives
\begin{equation}
\aligned
\ang{wv} &=0,\\
-\ang{wu}(z) &= -p_1z -\nu\aa^2U'''
\endaligned
\end{equation}
and leaves $\ang{w^2}$ undetermined up to an arbitrary function of $z$.
The identification in (\ref{SBM-alphavsRANS}) is exactly the same
(up to modified pressure and possibly $\ang{w^2}$) identification that
 was derived when identifying the NS-$\aa$ model (VCHE) with the
RANS equations  in the
 channel symmetry in \cite{CH98,CH99,CH00}.  The same identification holds
true in the case of the Leray-$\aa$ model \cite{CHOT}, the Clark-$\aa$
model in \cite{CHTi} and the ML-$\alpha$ model in \cite{ILT}.  Therefore, similar to the earlier alpha models, the general solution of simplified Bardina and NS-$\alpha$ will be identical (up to a modified pressure) and in particular,
 the mean flows in both cases are the same functions.  A similar result applies to turbulent pipe flows following the same argument and we will not include it here.  For further details regarding the identification of the equations under the pipe symmetry, see \cite{CH98,CH99,CH00,ILT}.

\section{Existence and Uniqueness} \label{E-U}
In this section we will prove the global existence and continuous dependence on initial data, (in particular,
the uniqueness of weak solution) of the system in (\ref{Bardina-func}).
 We will establish the estimates first for the finite dimensional Galerkin approximation scheme and then using the appropriate Aubin compactness theorems (see for, e.g.,
 \cite{CF88,TT84,TT88}) we can pass to the limit.  In this section, we fix $T > 0$ to be arbitrarily large.
\\\\
The finite dimensional Galerkin approximation, based on the eigenfunctions of the operator $A$, to (\ref{Bardina-func}) is :
\begin{equation}\label{Galerkin}
\begin{split}
\dfrac{d}{dt}(u_m + \alpha^2 Au_m) + \nu A(u_m+\alpha^2Au_m) + P_m B(u_m, u_m)&= P_mf\\
u_m(0)&=P_mu^{in}.
\end{split}
\end{equation}
\subsection{$H^1$ estimates}
We take the inner product of the Galerkin approximation (\ref{Galerkin}) with $u_m$ and use (\ref{buww}) to obtain
\begin{eqnarray}\label{h1}
\dfrac{1}{2}\dfrac{d}{dt}(|u_m|^2+\aa^2\|u_m\|^2) + \nu (\|u_m\|^2 + \aa^2|Au_m|^2)= (P_mf,u_m)=(f, P_mu_m) = (f, u_m).
\end{eqnarray}

Notice that by the Cauchy-Schwarz inequality, we have
\begin{eqnarray}
|(f,u_m)| \leq
\left\{
\begin{array}{ll}
|A^{-1}f||Au_m|\\
|A^{-1/2}f| \|u_m\|
\end{array} \right.
\end{eqnarray}
and by Young's inequality we have
\begin{eqnarray}
|(f,u_m)| \leq
\left\{
\begin{array}{ll}
\dfrac{|A^{-1}f|^2}{2\nu\aa^2}+\dfrac{\nu}{2}\aa^2 |Au_m|^2\\\\
\dfrac{|A^{-1/2}f|^2}{2\nu} + \dfrac{\nu}{2}\|u_m\|^2.
\end{array} \right.
\end{eqnarray}
We let $K_1 = \min \left\{\dfrac{|A^{-1/2}f|^2}{\nu}, \dfrac{|A^{-1}f|^2}{\nu\aa^2}\right\} $, from the above inequalities we get
\begin{eqnarray}\label{K1}
\dfrac{d}{dt}(|u_m|^2 + \aa^2\|u_m\|^2) + \nu(\|u_m\|^2 + \aa^2|Au_m|^2) \leq K_1.
\end{eqnarray}
Applying Poincar\'{e} inequality we get
\begin{eqnarray}
\dfrac{d}{dt}(|u_m|^2 + \aa^2\|u_m\|^2) + \nu\lambda_1(|u_m|^2 + \aa^2\|u_m\|^2) \leq K_1.
\end{eqnarray}
We then apply Gronwall's inequality to obtain
\begin{equation}\label{H1-ball}
\begin{split}
|u_m(t)|^2 + \aa^2\|u_m(t)\|^2 &\leq e^{-\nu\lambda_1 t}(|u_m(0)|^2+\aa^2\|u_m(0)\|^2)+\dfrac{K_1}{\nu\lambda_1}(1-e^{-\nu\lambda_1 t})
\end{split}
\end{equation}
 That is,
\begin{equation}\label{u_bound}
|u_m(t)|^2 + \aa^2\|u_m(t)\|^2 \leq k_1 :=|u^{in}|^2 + \aa^2\|u^{in}\|^2 + \dfrac{K_1}{\nu\lambda_1}
\end{equation}
Thus, for $t\in [0,T]$, where $T>0$ arbitrary but finite, we get  $u_m \in L^\infty([0,T],V)$, where the bound is uniform in $m$, provided $u^{in}\in V$.

\subsection{$H^2$ estimates}
Integrating (\ref{K1}) over the interval $(t,t+r) \mbox{ for } r > 0$, we obtain
\begin{equation}\label{H2-visc}
\aligned
\nu \int_t^{t+r}(\|u_m(s)\|^2 + \aa^2|Au_m(s)|^2) ds &\leq rK_1 + |u_m(t)|^2+\aa^2\|u_m(t)\|^2\\
&\leq rK_1 + k_1.
\endaligned
\end{equation}
Now, take the inner product of the Galerkin approximation (\ref{Galerkin}) with $Au_m$ to obtain
\begin{equation}
\dfrac{1}{2}\dfrac{d}{dt}(\|u_m\|^2+\aa^2|Au_m|^2)+\nu(|Au_m|^2+\aa^2|A^{3/2}u_m|^2) + (B(u_m,u_m),Au_m) = (f,Au_m).
\end{equation}
Notice that
\begin{equation}
|(f,Au_m)| \leq \left\{
\begin{array}{ll}
|A^{-1/2}f||A^{3/2}u_m|\\
|f||Au_m|.
\end{array} \right.
\end{equation}
Again by Young's inequality we have
\begin{eqnarray}
|(f,Au_m)| \leq \left\{
\begin{array}{ll}
\dfrac{|A^{-1/2}f|^2}{\nu\aa^2}+\dfrac{\nu}{4}\aa^2|A^{3/2}u_m|^2\\\\
\dfrac{|f|^2}{\nu}+\dfrac{\nu}{4}|Au_m|^2
\end{array} \right.
\end{eqnarray}
We denote by $K_2 = \min \left\{\dfrac{|A^{-1/2}f|^2}{\nu\aa^2},\dfrac{|f|^2}{\nu}\right\}$.  Then we have
\begin{eqnarray}\label{K2}
\dfrac{1}{2}\dfrac{d}{dt}(\|u_m\|^2 + \aa^2|Au_m|^2) + \dfrac{3\nu}{4}(|Au_m|^2+\aa^2|A^{3/2}u_m|^2)\leq K_2 + |(B(u_m,u_m),Au_m)|.
\end{eqnarray}
Using H\"older inequality, (\ref{sobolev}) and Young's inequality
\begin{eqnarray}
\aligned
|(B(u_m,u_m),Au_m)|
&\leq c \|u_m\|\|u_m\|^{1/2}|Au_m|^{1/2}|Au_m|\\
& = c\|u_m\|^{3/2}|Au_m|^{3/2} \\
&\leq \dfrac{\nu}{4}|Au_m|^2 + c\|u_m\|^{6}.
\endaligned
\end{eqnarray}
Using the above estimates and (\ref{K2}) we obtain
\begin{eqnarray}
\dfrac{d}{dt}(\|u_m\|^2 + \aa^2|Au_m|^2)+\nu(|Au_m|^2+\aa^2|A^{3/2}u_m|^2) \leq 2 K_2 + c\|u_m\|^6.
 \end{eqnarray}
We integrate the above equation over the interval $(s,t)$ and use (\ref{u_bound}) and (\ref{H2-visc}) to obtain:
\begin{eqnarray}\label{int}
\|u_m(t)\|^2+\aa^2|Au_m(t)|^2 \leq \|u_m(s)\|^2+\aa^2|Au_m(s)|^2 + 2(t-s)K_2 + c\biggl(\dfrac{k_1}{\aa^2}\biggr)^3(t-s)
\end{eqnarray}
Now, we integrate with respect to $s$ over the interval $(0,t)$ and use (\ref{H2-visc})
\begin{eqnarray}\label{shorttime}
t(\|u_m(t)\|^2+\aa^2|Au_m(t)|^2)\leq\dfrac{1}{\nu}(tK_1+k_1)+t^2K_2 + c \biggl(\dfrac{k_1}{\aa^2}\biggr)^3 \dfrac{t^2}{2}
\end{eqnarray}
for all $t\geq 0$.\\
For $t\geq\frac{1}{\nu\lambda_1}$ we integrate (\ref{int}) with respect
to $s$ over the interval $(t-\frac{1}{\nu\lambda_1},t)$
\begin{equation}\label{largetime}
\aligned
&\dfrac{1}{\nu\lambda_1}(\|u_m(t)\|^2+\aa^2|Au_m(t)|^2)\\&
\leq
\dfrac{1}{\nu}\left(\dfrac{1}{\nu\lambda_1}K_1 + k_1\right)+
 2 K^2 \biggl(\dfrac{1}{2\nu\lambda_1}\biggr)^2 + c\biggl(\dfrac{k_1}{\aa^2}\biggr)^3 \biggl(\dfrac{1}{2\nu\lambda_1}\biggr)^2
\endaligned
\end{equation}

Thus, from (\ref{shorttime}) and (\ref{largetime}) we conclude:
\begin{equation}\label{total}
\|u_m(t)\|^2+\aa^2|Au_m(t)|^2 \leq k_2(t)
\end{equation}
for all $t>0$.  We note that, $k_2(t)$ enjoys the following properties:
\begin{enumerate}
\item  $k_2(t)$ is finite for all $t > 0$.
\item  If $u^{in}\in V$, but $u^{in}\notin D(A)$, then the $\lim_{t\rightarrow 0^{+}}k_2(t) = \infty$.
\item $\limsup_{t\rightarrow \infty}k_2(t) < \infty$.
\end{enumerate}
\begin{remark}
>From (\ref{int}), one can observe that if $u^{in}\in D(A)$, then $u_m(\cdot)$ is bounded uniformly in the
$L^\infty([0,T];D(A))$~norm, independently of $m$.  On the other hand, if $u^{in}\in V$, but $u^{in}\not\in D(A)$,
 we conclude from the above that $u_m \in L^\infty_{loc}((0,T],D(A))\cap L^2([0,T],D(A))$.
\end{remark}
In order to extract convergent subsequence by using Aubin's lemma (see \cite{CF88,Lions,TT84}),
we need to establish estimates for $\dfrac{dv_m}{dt},\dfrac{du_m}{dt}$.
\begin{equation}
\dfrac{dv_m}{dt} = -\nu Av_m -B(u_m,u_m) + P_mf
\end{equation}
Take the $D(A)'$ (the dual of the space $D(A)$) action of the
equation above with $w\in D(A)$, we observe that
\begin{equation}
|(P_mf,w)| = |(f,P_mw)|\leq |A^{-1}f||Aw|\leq \lambda_1^{-1}|f||Aw| = L^2|f||Aw|
\end{equation}
and using (\ref{buvw}), we have
\begin{equation}\label{P_m}
\begin{split}
|(P_mB(u_m,u_m),w)| &\leq c |u_m|^{1/2}\|u_m\|^{1/2}\|u_m\|\|w\|\\
&= c|u_m|^{1/2}\|u_m\|^{3/2}\|w\| \\
&= c\lambda_1^{-1/2}|u_m|^{1/2}\|u_m\|^{3/2}|Aw|.
\end{split}
\end{equation}
By (\ref{u_bound}), $\|u_m\|_{L^\infty([0,T];V)}$ is bounded uniformly with respect to $m$.
Thus by (\ref{P_m}), we can deduce that $\|P_mB(u_m,u_m)\|_{L^2([0,T];D(A)')}$~is also bounded uniformly with respect to $m$.  Now, the uniform in $m$, $L^2([0,T];D(A))$
 bound for $u_m$ implies that $\|v_m\|_{L^2([0,T];H)}$ is uniformly bounded, which in turn implies that $\|Av_m\|_{L^2([0,T];D(A)')}$ is uniformly bounded, as well. Thus, we conclude, $\|\dfrac{dv_m}{dt}\|_{L^2([0,T];D(A)')}$, and in particular, $\|\dfrac{du_m}{dt}\|_{L^2([0,T];H)}$, are uniformly bounded with respect to $m$.
By Aubin compactness theorem (see, e.g., \cite{CF88,Lions, TT84})
we conclude that there is a subsequence $u_{m'}(t)$ and a function $u(t)$ such that
\begin{equation}\label{conv-u}
\begin{split}
&u_{m'}(t) \rightarrow u(t) \mbox{ weakly in } L^2([0,T];D(A)) \\
&u_{m'}(t) \rightarrow u(t) \mbox{ strongly in } L^2([0,T];V) \\
&u_{m'}\rightarrow u \mbox{ in } C([0,T];H),
\end{split}
\end{equation}
or equivalently,
\begin{equation}\label{v-conv}
\begin{split}
&v_{m'}(t) \rightarrow v(t) \mbox{ weakly in } L^2([0,T];H) \\
&v_{m'}(t) \rightarrow v(t) \mbox{ strongly in } L^2([0,T];V') \\
&v_{m'}\rightarrow v \mbox{ in } C([0,T];D(A)').
\end{split}
\end{equation}
We relabel $u_{m'}$ and $v_{m'}$ with $u_m$ and $v_m$ respectively.  Let $w \in D(A)$, then we have
\begin{equation}
(v_m(t), w) + \nu\int_{t_0}^t(v_m(s),Aw) ds+\int_{t_0}^t(B(u_m(s), u_m(s)), P_mw) ds = (v_m(t_0), w) + (f, P_mw)(t-t_0)
\end{equation}
for all $t_0, t\in [0,T]$.  The sequence $v_m(t)$ converges weakly in $L^2([0,T];H)$ and thus,
\begin{equation}
\lim_{m\rightarrow\infty}\int_{t_0}^t(v_m(s),Aw)ds = \int_{t_0}^t(v(s),Aw)ds.
\end{equation}
Also, by (\ref{v-conv}), $v_m(t)$ converging weakly in $L^2([0,T];H)$ implies that there is a subsequence of $v_m$,  which we relabel as $v_m$, which converges a.e. $t\in [0,T]$ to $v(t)$ in $H'\simeq H$.  Thus, we conclude that
\begin{equation}
\begin{split}
(v_m(t),w) & \rightarrow (v(t),w), \mbox{ and }\\
(v_m(t_0),w) & \rightarrow (v(t_0),w)
\end{split}
\end{equation}
for a.e. $t,t_0 \in [0,T]$.
On the other hand,
\begin{equation}
\Bigl\vert\int_{t_0}^t(B(u_m(s),u_m(s)), P_m w) - \left(B(u(s),u(s)),w\right)ds\Bigr\vert\le I_m^{(1)} + I_m^{(2)} + I_m^{(3)}.
\end{equation}
Using (\ref{buvw2}) and Agmon's inequality (\ref{Agmon}), we get
\begin{equation}
\begin{split}
I_m^{(1)} &= \Bigl\vert\int_{t_0}^t\left(B(u_m(s),u_m(s)),P_mw-w\right)ds\Bigr\vert\\
&\leq c \int_{t_0}^t |u_m(s)|\|u_m(s)\|\|P_mw-w\|_{L^\infty(\Omega)}ds\\
&\leq c \biggl(\int_{t_0}^t|u_m(s)|^2 ds\biggr)^{1/2}\biggl(\int_{t_0}^t{\|u_m(s)\|^2ds}\biggr)^{1/2}|P_mw - w|^{1/4}|A(P_mw-w)|^{3/4}.
\end{split}
\end{equation}
Since $u_m$ is bounded uniformly in $L^\infty([0,T];V)$ independent of m , and thus is bounded uniformly
in $L^\infty([0,T];H)$ thanks to Poincar\'{e} inequality, we get $\lim_{m\rightarrow\infty}I_m^{(1)}=0$.
\\\\
Again, using (\ref{buvw2}), Agmon's inequality (\ref{Agmon}), and Poincar\'{e} inequality, we get
\begin{equation}
\begin{split}
I_m^{(2)} &= \Bigl\vert\int_{t_0}^t\left(B(u_m(s)-u(s),u_m(s)),w\right)ds\Bigr\vert\\
&\leq c \int_{t_0}^t |u_m(s)-u_m(s)|\|u_m(s)\|\|w\|_{L^\infty(\Omega)}ds\\
&\leq c \biggl(\int_{t_0}^t|u_m(s)-u(s)|^2ds\biggr)^{1/2}\biggl(\int_{t_0}^t{\|u_m(s)\|^2ds}\biggr)^{1/2}\lambda^{-1/4}|Aw|.
\end{split}
\end{equation}
Now since $u_m \rightarrow u$ strongly in $L^2([0,T];V)$ (thus in $L^2([0,T];H)$  ) and $u_m$ is bounded uniformly independent of $m$
in $L^\infty([0,T];V)$ , we get that $\lim_{m\rightarrow\infty}I_m^{(2)}=0$.
\begin{equation}
\begin{split}
I_m^{(3)}&= \Bigl\vert\int_{t_0}^t\left(B(u(s),u_m(s)-u(s)),w\right)ds\Bigr\vert\\
&\leq c\int_{t_0}^t |u(s)|\|u_m(s)-u(s)\|\|w\|_{L^\infty(\Omega)}ds\\
&\leq c \biggl(\int_{t_0}^t|u(s)|^2ds\biggr)^{1/2}\biggl(\int_{t_0}^t \|u_m(s)-u(s)\|^2ds\biggr)^{1/2}\lambda_1^{-1/4}|Aw|
\end{split}
\end{equation}
where we applied (\ref{buvw}) and (\ref{buvw2}) in the second and third inequality, respectively.
Now, $u_m\rightarrow u$ strongly in $L^2([0,T];V)$ implies that $\lim_{m\rightarrow\infty}I_m^{(3)}=0$.\\
>From the above calculations, we have that for a.e. $t_0, t \in [0,T]$,
\begin{equation}\label{end-result}
(v(t), w)-(v(t_0),w) + \nu\int_{t_0}^t(v,Aw)ds+\int_{t_0}^t \left(B(u(s),u(s)),w\right)ds = \int_{t_0}^t(f,w)ds.
\end{equation}
for every $w \in D(A)$.  To show that $v\in C([0,T];V')$, and hence, $u\in C([0,T];V)$, we want to show that the viscous
term $\nu \int_{t_0}^t (v(s), Aw) ds$ and the nonlinear term $\int_{t_0}^t \ang{B(u,u), w}_{D(A)'}ds \rightarrow 0$ as $t\rightarrow t_0$.
\begin{equation}
\begin{split}
\Bigl\vert\nu \int_{t_0}^t (v(s),Aw) ds \Bigr\vert &\leq \nu \left(\int_{t_0}^t|v(s)|^2ds\right)^{1/2}\left(\int_{t_0}^t |Aw|^2ds\right)^{1/2}
\rightarrow 0 \quad \mbox{ as } \quad t\rightarrow t_0,
\end{split}
\end{equation}
since $v\in L^2([0,T];H)$ and $w\in D(A)$.
\begin{equation}
\Bigl\vert\int_{t_0}^t \left((B(u(s),u(s)),w)ds\right)\Bigr\vert \leq |w|_{L^\infty(\Omega)}\left(\int_{t_0}^t|u(s)|^2ds\right)^{1/2}
\left(\int_{t_0}^t\|u(s)\|^2ds\right)^{1/2} \rightarrow 0 \quad \mbox{ as } \quad t\rightarrow t_0,
\end{equation}
since $u\in L^\infty([0,T];V)$.  Thus, this implies that for a.e. $t\in[0,T]$, $(v(t),w)\rightarrow (v(t_0),w)$
as $t\rightarrow t_0$, for every $w\in D(A)$.  In particular, $v(t)\in H \subset V'$ and $w\in D(A)\subset V$,
implies that for a.e. $t\in[0,T]$, $\ang{v(t),w}_{V'}\rightarrow \ang{v(t_0),w}_{V'}$ as $t\rightarrow t_0$,
 for every $w\in D(A)$.  Since $D(A)$ is dense in $V$,
 for any test function $\phi \in V$ and for
 every $\epsilon > 0$, there exists a
$w\in D(A)$ such that $\|w-\phi \|< \epsilon/(M+1)$, where $M =
2\sup_{t\in [0,T]}\|v(t)\|_{V'}$.  Thus for every $\phi \in V$
\begin{equation}
|\ang{v(t)-v(t_0),\phi}_{V'}|\leq |\ang{v(t)-v(t_0),w}_{V'}|+|\ang{v(t)-v(t_0),w-\phi}_{V'}|.
\end{equation}
The first term goes to zero as $t\rightarrow t_0$ since $w\in D(A)$.  For the second term, we have
\begin{equation}
|\ang{v(t)-v(t_0),w-\phi}_{V'}| \leq \|v(t)-v(t_0)\|_{V'} \|w-\phi\|
\leq M \|w-\phi\| < \epsilon.
\end{equation}
Since $\epsilon>0$ is arbitrary, we conclude that
$\ang{v(t)-v(t_0),\phi}_{V'}\rightarrow 0$, as $t\rightarrow t_0$,
for all $\phi\in V$. Hence, $v\in C([0,T];V')$ and in particular,
$u\in C([0,T];V)$.
\\\\
To summarize: we have established above the global existence of weak solution of the simplified Bardina system
by the standard Galerkin approximation scheme together with some useful {\it a priori} estimates.

\begin{theorem}{\bf(Global existence and uniqueness)} Let $f \in H$ and $u^{in} \in V$.  Then for any $T > 0$, (\ref{Bardina-func}) has a unique weak solution u in \mbox{$[0,T]$}.
\end{theorem}
To complete the proof of the theorem above, we are left to prove the uniqueness of weak solutions.
\subsection*{Uniqueness of Weak Solution}
Next we will show the continuous dependence of the weak solutions
 in the appropriate norm specified below, on the initial data, and in particular, the uniqueness of weak solutions.

Let $u$ and $\ubar$ be any two weak solutions of (\ref{Bardina-func}) on the interval
$[0,T]$, with initial values $u(0) = u^{in} \in V$ and
$\ubar(0) = \ubar^{in} \in V$, respectively.  Let us denote by
$v = (u + \aa^2Au)$, $\vbar = (\ubar + \aa^2A\ubar)$, $\dd u = u - \ubar$,
and by $\dd v = v - \vbar$.  Then from (\ref{Bardina-func}) we get:
\begin{equation}
\dfrac{d}{dt}\dd v + \nu A\delta v+ B(\dd u, u) + B(\ubar, \dd u)=0 \label{U2}
\end{equation}
By taking the $D(A)'$ action of (\ref{U2}) with $\dd u$,
\begin{equation}
\ang{\dfrac{d}{dt}\dd v, \dd u}_{D(A)'} + \nu\ang{A\delta v, \delta u}_{D(A)'}+\left(B(\dd u, u), \dd u\right) + \left(B(\ubar, \dd u),\dd u\right) = 0
\end{equation}
and by applying a Lemma of Lions-Magenes concerning the derivative of functions with
values in Banach space, (cf. Chap. III-p.169-\cite{TT84}) and by the property of the bilinear form $B$, (\ref{buww}), we get:
\begin{equation}
\dfrac{1}{2}\dfrac{d}{dt}(|\dd u|^2 + \aa^2\|\dd u\|^2)+ \nu (\|\delta u\|^2+\aa^2|A\delta u|^2)+\left(B(\dd u,u),\dd u\right)=0.
\end{equation}
Dropping the nonnegative viscous term, and by using property (\ref{buvw}), we get
\begin{equation}
\aligned
\dfrac{1}{2}\dfrac{d}{dt}(|\dd u|^2 + \aa^2\|\dd u\|^2) &\leq c\ \|\dd u\|^{2}|u|^{1/2}\|u\|^{1/2}\\
& \leq c\lambda_1^{-1/4}\|\dd u\|^2\|u\|\\
\endaligned
\end{equation}
By Gronwall inequality, we obtain:
\begin{equation}\label{cont-dep}
(|\dd u(t)|^2 + \aa^2\|\dd u(t)\|^2) \leq (|\dd u(0)|^2 +
\aa^2\|\dd u(0)\|^2)\exp
\left(\int_0^t
\dfrac{C\|u(s)\|}{\aa^2}ds\right).
\end{equation}
In (\ref{cont-dep}), since $u\in L^\infty([0,T];V)$, we have shown
the continuous dependence of the weak solutions on the initial data in the $L^\infty([0,T];V)$ norm.  
In particular, in the case we have the same initial data, we have  $\|\dd u(t)\|^2 = 0$, which implies that we have $u(t) = \bar{u}(t)$, for all $t\in[0,T]$.
\section{Global Attractors, Their Dimensions and
Connection to Dissipation Length Scales}   \label{S-A}

Now that we have established the global well-posedness to the simplified Bardina model, in this section we will show the existence of global attractor
 $\mathcal{A} \subset  V$ for the system (\ref{Bardina-func}), its finite Hausdorff and fractal dimensions, and the physical relevance of this
finite dimension of global attractor to the concept of ``finite dimensionality'' of turbulent flows.

Following standard techniques, the method that we will use to estimate the dimension of the global attractor stems from the following lemmas (see \cite{CF88,LT76,TT88} and \cite{FHTM}, respectively):
\begin{lemma}\label{lemma4}{\bf(The Lieb-Thirring inequality).}
 Let $\lbrace\psi_j\rbrace_{j=1}^N$ be an orthonormal set of functions
 in $(H)^k = \underbrace{H \oplus H \dots \oplus H}_{\mbox{k-times}}$.
  Then there exists a constant $C_{LT}$, which depends on $k$, but is
  independent of $N$, such that
\begin{equation}\label{LT}
\int_\Omega\left(\sum_{j=1}^N \psi_j(x)\cdot\psi_j(x)\right)^{5/3} dx
\leq C_{LT} \sum_{j=1}^N \int_\Omega (\nabla\psi_j(x):\nabla\psi_j(x)) dx.
\end{equation}
\end{lemma}

\begin{lemma}\label{lemma5}
 Let $\lbrace\phi_j\rbrace_{j=1}^N\in V$ be an orthonormal set of
functions  with respect to the inner product $[\cdot,\cdot]$:
$$
[\phi_i,\phi_j]=(\phi_i,\phi_j)+\alpha^2((\phi_i,\phi_j))=
\delta_{ij}.
$$
Let $\psi_j(x)=(\phi_j(x), \alpha\frac{\partial\phi_j(x)}{\partial x_1},
\alpha\frac{\partial\phi_j(x)}{\partial x_2},
\alpha\frac{\partial\phi_j(x)}{\partial x_3})$,
and
$\phi^2(x)=\sum_{j=1}^N(\phi_j(x)\cdot\phi_j(x))$.
  Then there exists a constant $C_{F}$, which is
  independent of $N$, such that
\begin{equation}\label{infty}
\|\phi\|_{L^\infty}^2\le
\frac{C_F}{\alpha^2}
\left(\sum_{j=1}^N\int_{\Omega}(\nabla\psi_j(x):\nabla\psi_j(x))dx
\right)^{1/2}.
\end{equation}
\end{lemma}

To start the study of the finite dimensionality of the Hausdorff and fractal dimensions of the global attractor, first we recall that, from the existence and uniqueness properties of the solutions
to (\ref{Bardina-func}), we get a semi-group of solution operators, denoted
as $\lbrace S(t)\rbrace_{t\geq 0}$, which associates, to each
$u^{in} \in V$, the semi-flow for time $t\geq 0: S(t)u^{in} = u(t)$.
We are now ready state and prove the following theorem:
\begin{theorem}
There is a compact global attractor $\mathcal{A}\subset V$, in terms of the solution $u$, for the system (\ref{Bardina-func}).  Moreover, we have an upper bound for the
Hausdorff and fractal dimension of the attractor $\mathcal{A}$
\begin{equation}
d_H(\mathcal{A})\leq d_F(\mathcal{A})\leq c\
G^{6/5}\left(\dfrac{1}{\lambda_1^{9/5}\aa^{18/5}}\right)= c\ G^{6/5}\left(\dfrac{L}{\aa}\right)^{18/5}
\end{equation}
where $G = \dfrac{|f|}{\nu^2\lambda_1^{3/4}}$ is the Grashoff number.
\end{theorem}
\begin{proof}
The first requirement to show the existence of the nonempty compact attractor is to show that we have an absorbing ball in $V$ and $D(A)$ and that the semigroup $\lbrace S(t)\rbrace_{t\ge 0}$ defined above is compact (see, e.g., \cite{CF88, JH88, JR11, SY11, TT88}). This can be established from the previous {\it a priori} estimates. First, let us show that there is an absorbing ball in $V$ and $D(A)$.
By (\ref {H1-ball}),(\ref{conv-u}), and the fact that $|u_m(0)|\leq |u(0)| $ and $\|u_m(0)\|\leq \|u(0)\| $ we have, by passing to the limit with $m\rightarrow\infty$,
\begin{equation}
|u(t)|^2 + \aa^2\|u(t)\|^2 \leq
e^{-\nu\lambda_1t}(|u(0)|^2+\aa^2\|u(0)\|^2)+
\dfrac{K_1}{\nu\lambda_1}(1-e^{-\nu\lambda_1t}).
\end{equation}
Choose $t$ large enough such that
$e^{-\nu\lambda_1t}(|u(0)|^2+\aa^2\|u(0)\|^2)\le\dfrac{K_1}{\nu\lambda_1}$,
 then we have
\begin{equation}\label{R_v}
|u(t)|^2 + \aa^2\|u(t)\|^2 \leq 2\dfrac{K_1}{\nu\lambda_1},
\end{equation}
where we recall
$K_1 = \min \left\{ \dfrac{|A^{-1}f|^2}{\nu \aa^2},
\dfrac{|A^{-1/2}f|^2}{\nu} \right\} $.  In particular,
\begin{equation}
\limsup_{t \rightarrow \infty}\ (|u(t)|^2 + \aa^2\|u(t)\|^2)
\leq 2\dfrac{K_1}{\nu\lambda_1} =: R_V^2.
\end{equation}
Therefore, the system (\ref{DLE}) has the ball $\mathcal{B}_V(0)$ in $V$
of radius $R_V$ as an absorbing ball in V.\\
Proving the existence of absorbing ball $\mathcal{B}_{D(A)}(0)$ in $D(A)$ is similar.  By (\ref{largetime}) and
(\ref{total}) we conclude that
\begin{equation}
\limsup_{t \rightarrow \infty}\ (\|u(t)\|^2+\aa^2|Au(t)|^2)
\leq \limsup_{t \rightarrow \infty} k_2(t) =: R_{D(A)}^2 < \infty,
\end{equation}
and therefore we have the ball $\mathcal{B}_{D(A)}(0)$  in $D(A)$ with
radius $R_{D(A)}$ as an absorbing ball in D(A).

Now applying Rellich lemma \cite{AR75} we have that
$S(t): V\rightarrow \mathcal{D}(A) \subset\subset V$, for $t>0$, is a
compact semigroup from $V$ to itself.  What is left is to show that indeed we have a nonempty compact attractor.  Since $S(t)\mathcal{B}_V(0) \subset \mathcal{B}_V(0)$, it follows that
for each $s> 0$ the set
$C_s:=\overline{\cup_{t\ge s}S(t)\mathcal{B}_V(0)}^V$ is nonempty and
compact in $V$.  By monotonicity of  $C_s$ for $s> 0$, and by the finite
intersection property of compact sets, we see that
\begin{equation}
\mathcal{A} = \bigcap_{s>0}
\overline{\bigcup_{t\ge s}S(t)\mathcal{B}_V(0)}^V\subset V
\end{equation}
is a nonempty compact set in $V$ and indeed is the unique global
attractor in $V$.

We are now ready to give an upper bound estimate to the Hausdorff and fractal dimensions of the global attractor.  As mentioned above, we will use the trace formula (see, e.g., \cite{CF85,CF88,TT88}) to establish this estimate.

  The first step in this estimation is to do linearization about a solution.
  We note that in order to apply the techniques in \cite{CF85,CF88,FHTM,TT88},
we need that the mapping $S(t):V\rightarrow V$ is differentiable with respect to initial data.
Following similar ideas of energy estimates in the proof of uniqueness of weak solutions in the previous section,
one can show that $S(t)u^{in}$ is differentiable with respect to $u^{in}$, when $u^{in} \in\mathcal{A}$.  Thus said, we
linearize the viscous simplified Bardina model (\ref{Bardina-func}) about a
solution $u(t)$ (or $v(t) = u(t) + \aa^2 Au(t)$)
\begin{equation}\label{perturb-v}
\begin{split}
\dfrac{d}{dt}\dd v + \nu A\dd v + B(\dd u, u)  + B(u, \dd u) &= 0\\
\delta v(0) = \delta v^{in} &= \delta u^{in} + \aa^2 A \delta u^{in}
\end{split}
\end{equation}
where $\dd v$ is a perturbation satisfying (\ref{perturb-v})  and is given by $\dd v = \dd u + \aa^2 A \dd u$.  With this relationship, $\dd u$ evolves
according to the equation
\begin{equation} \label{evol_u}
\begin{split}
\dfrac{d}{dt}\dd u + \nu A \dd u +
 (I + \aa^2 A)^{-1}[B(\dd u ,u)+ B(u, \dd u)] &= 0\\
\delta u(0) &= \delta u^{in},
\end{split}
\end{equation}
which we write symbolically as
\begin{equation}
\begin{split}
\dfrac{d}{dt}\dd u + T(t) \dd u &= 0\\
\delta u(0) &= \delta u^{in},
\end{split}
\end{equation}
where $T(t)\psi = \nu A\psi +
(I + \aa^2A)^{-1}[B(\psi,u(t))+ B(u(t),\psi)]$.
Let $\dd u_i(0)$, $j = 1,\dots, N$ be a set of linearly independent
vectors in $V$ and let $\dd u_j(t)$ be the corresponding solutions of
 (\ref{evol_u}) with initial value $\dd u_j(0)$~for~$j = 1, \dots, N$.
  Let
\begin{equation}\label{Trace}
\mathcal{T}_N(t) = \mbox {Trace}(P_N(t)\circ T(t)\circ P_N(t))
\end{equation}
where $P_N(t)$ is the orthogonal projection of $V$ onto the span
$\lbrace\dd v_1(t),\dd v_2(t),\dots,\dd v_N(t)\rbrace$.
We shall denote by $\lbrace\phi_j(t)\rbrace_{j=1,\dots,N}$, an orthonormal basis,
with respect to inner product
$[\cdot,\cdot]= (\cdot,\cdot) + \aa^2 ((\cdot,\cdot))$
of the space $P_N V=\mbox{span}\{\delta v_1(t),\dots,\delta v_2(t)\}$.
>From (\ref{Trace}) we have
\begin{equation}
\aligned
\mathcal{T}_N(t)&=\sum_{j=1}^N\ [T(t)\phi_j(\cdot,t),\phi_j(\cdot,t)]\\
&=\sum_{j = 1}^N \nu [A\phi_j,\phi_j]+[(I+\aa^2A)^{-1}
B((\phi_j,u),\phi_j]+[(I+\aa^2A)^{-1}B(u,\phi_j),\phi_j]\\
&=\nu \sum_{j=1}^N\ [A\phi_j,\phi_j]+\sum_{j=1}^N\
 (B(\phi_j,u),\phi_j)
+\sum_{j=1}^N\ (B(u,\phi_j),\phi_j)\\
&=\nu \sum_{j=1}^N\ [A\phi_j,\phi_j]+\sum_{j=1}^N\
 (B(\phi_j,u),\phi_j)
\endaligned
\end{equation}
By the definition of the inner product
$[\cdot,\cdot]$, we have
\begin{equation}\label{T=Q+R}
\aligned
 \sum_{j=1}^N\ [A\phi_j,\phi_j]=
 \sum_{j=1}^N\ (A\phi_j,\phi_j) + \aa^2 \sum_{j=1}^N(A\phi_j,A\phi_j)=
 \sum_{j=1}^N\int_\Omega(\nabla\psi_j(x,t):\nabla\psi_j(x,t))dx
 =:Q_N(t)
 \endaligned
\end{equation}
where,
\begin{equation}\label{psi_j}
\psi_j = \left(\phi_j,\ \aa\dfrac{\pp}{\pp x_1}\phi_j,\ \aa\dfrac{\pp}{\pp x_2}\phi_j,\ \aa\dfrac{\pp}{\pp x_3}\phi_j   \right)^T.
\end{equation}
Note also that
\begin{equation}\label{dirac}
(\psi_j,\psi_k)=\dd_{jk}.
\end{equation}
Setting
$$
\mathcal{R}_N(t)=\sum_{j=1}^N(B(\phi_j,u),\phi_j),
$$
we have
\begin{equation}\label{trace}
\mathcal{T}_N(t)=\nu Q_N(t)+\mathcal{R}_N(t).
\end{equation}
We denote by $\psi^2:=\sum_{j=1}^N \psi_j \cdot \psi_j$.
For $\mathcal{R}_N(t)$ we have
\begin{equation}\label{R_N1_bound}
\begin{aligned}
|\mathcal{R}_N(t)|\leq \sum_{j=1}^N |(B(\phi_j,u),\phi_j)|
\leq \int_\Omega\sum_{j=1}^N|(\phi_j\cdot\nabla)u\ \phi_j|dx
\leq  \int_\Omega\sum_{j=1}^N\phi_j^2\ |\nabla u|dx\\
\leq \dfrac{C_F}{\aa^2}Q_N^{1/2} \biggl(\int_\Omega |\nabla u|^2dx\biggr)^{1/2}\biggl(\int_\Omega 1 dx\biggr)^{1/2}=\dfrac{C_F|\Omega|^{1/2}}{\aa^2}Q_N^{1/2}\|u(t)\|
\leq \dfrac{\nu}{2}Q_N + \dfrac{C_F^2|\Omega|}{2\nu\aa^4}\|u(t)\|^2
\end{aligned}
\end{equation}

By the estimates so obtained above we finally find
\begin{equation}\label{est-for-trace}
\mathcal{T}_N(t) \ge \dfrac{\nu}{2}Q_N(t)-\dfrac{C_F^2| \Omega|}{2\nu\aa^4} \|u(t)\|^2.
\end{equation}

By the asymptotic behavior of the eigenvalues of the operator $A$
(see (\ref{asymptotic})) and (\ref{dirac}) we get
\begin{equation}\label{Q}
Q_N(t) = \sum_{j=1}^N \|\psi_j\|^2 \geq
 \sum_{j=1}^N \lambda_j \geq c_0 \lambda_1 N^{5/3}.
\end{equation}
Now, by the trace formula (see, e.g., \cite{CF85, CF88,TT88}
and the references therein)
 if $N$ is large enough so that
\begin{equation}\label{tracethm}
\liminf_{T\rightarrow \infty}\dfrac{1}{T}\int_0^T \mathcal{T}_N(t)\ dt > 0
\end{equation}
then $N$ is an upper bound for the Hausdorff and fractal dimensions \cite{CF88,TT88}, (see also \cite{CI}), of the global attractor.\\
Thus, by (\ref{est-for-trace}) and (\ref{Q}) it is sufficient to
require $N$ to be large enough such that
\begin{equation}\label{ineq}
\nu\lambda_1N^{5/3} >
 \sup_{u^{in}\in\mathcal{A}}
\limsup_{T\rightarrow\infty}
\dfrac{1}{T}\int_0^T \dfrac{C_F^2 | \Omega| }{2\nu\aa^4}\|u(t)\|^2 dt.
\end{equation}

On the other hand, using H\"{o}lder inequality
we get from~(\ref{H2-visc})
\begin{equation}\label{time-average}
\begin{aligned}
\limsup_{T\rightarrow\infty}\dfrac{1}{T}\int_0^T\dfrac{C_F^2| \Omega|}{2\nu\aa^2}\|u(t)\|^2 dt\leq \dfrac{C_F^2 | \Omega|}{2\nu\aa^4}\cdot\dfrac{|f|^2}{\aa^2\nu^2\lambda_1}
\end{aligned}
\end{equation}
which implies
\begin{equation}
N^{5/3}\geq \dfrac{C| \Omega||f|^2}{\nu^4\lambda_1^3\aa^6}=\dfrac{|f|^2}{\nu^4\lambda_1^{3/2}}\cdot \dfrac{C| \Omega|}{\lambda_1^{3/2}\aa^6}\geq G^2\cdot\dfrac{C| \Omega|}{\lambda_1^{3/2}\aa^6}
\end{equation}
and we recall that $| \Omega| = (2\pi L)^3$ and that $\lambda_1 = L^{-2}$, thus,
\begin{equation}
N\geq G^{6/5} \dfrac{C}{\lambda_1^{9/5}\aa^{18/5}}
\end{equation}
>From this we deduce that
\begin{equation}
d_H(\mathcal{A})\leq d_F(\mathcal{A})\leq G^{6/5} \dfrac{C}{\lambda_1^{9/5}\aa^{18/5}}.
\end{equation}
\end{proof}


The interpretation of the upper bound estimate that we get for the Hausdorff and fractal dimension of the global attractor in terms of small scales is important in showing the finite dimensionality of flows and in particular in showing the numerical computability of the turbulence model.  To do this, we interpret the estimate for the attractor dimension in terms of the mean rate of energy dissipation of the simplified Bardina model. Following {\bf \cite{FHTM}} we define the corresponding mean rate of dissipation of ``energy'' for the simplified Bardina model (see (\ref{h1})) as
\begin{equation}\label{diss}
\bar{\epsilon} = L^{-3}\nu\sup_{u^{in}\in\mathcal{A}}
\limsup_{T\rightarrow\infty}\dfrac{1}{T}
\int_0^T (\|u(s)\|^2+\aa^2|Au(s)|^2)ds.
\end{equation}
Thus, and in analogy with the Kolmogorov dissipation length in the
classical theory of turbulence, we set the dissipation length scale for
the simplified Bardina model as
\begin{equation}\label{ld}
l_d = \left(\frac{\nu^3}{\bar{\epsilon}}\right)^{1/4}.
\end{equation}
Identifying the dimension of global attractor with the number of
degrees of freedom, we will show that the number of degrees of freedom
for the simplified Bardina model is bounded from above by a quantity which scales
like $(L/\aa)^{12/5}(L/l_d)^{12/5}$.

In fact, in view of~(\ref{diss}) we can write~(\ref{time-average})
as follows
\begin{equation}
\begin{aligned}
\limsup_{T\rightarrow\infty}\dfrac{1}{T}\int_0^T\dfrac{C_F^2| \Omega|}{2\nu\aa^2}\|u(t)\|^2 dt\leq\limsup_{T\rightarrow\infty}\dfrac{1}{T}\int_0^T\dfrac{C_F^2| \Omega|}{2\nu\aa^2}\left(\|u(t)\|^2+\aa^2|Au(t)|^2\right) dt
= \dfrac{C_F^2 L^6}{2\nu^2\aa^2}\bar{\epsilon} = \dfrac{c L^6 \bar{\epsilon}}{\nu^2\aa^4}
\end{aligned}
\end{equation}
Using this in~(\ref{ineq}) and recalling~(\ref{ld})
we obtain the following estimate for
 the dimension of the global attractor
 and hence the upper bound
 on the number of degrees of freedom in the simplified Bardina model is:
 \begin{equation}
d_H(\mathcal{A})\leq d_F(\mathcal{A})\leq c
\left(\dfrac{L}{\aa}\right)^{12/5}\left(\dfrac{L}{l_d}\right)^{12/5}.
\end{equation}
We remark that following the recent work of \cite{Holm-Gibbon} one can also interpret this bound in terms of the Reynolds number.
\section{Energy Spectra}\label{ES}
Turbulent flows are characterized by the presence of wide range of eddy sizes starting from the size of the flow domain, say $2\pi L$ in our case,  to much smaller scales, which become progressively smaller relative to $2\pi L$ as we increase the Reynolds number.  It is important to examine how the energy in a turbulent flow is distributed among these different size eddies by considering the energy spectrum.
Following similar arguments to those presented in \cite{FOIAS} and
 \cite{FHTP} (see also \cite{CHOT,CHTi,FMRT,ILT}) we will study in this section
the energy spectra of the simplified Bardina model. We will obtain
our results about the decay of the energy spectrum for the filtered
velocity $u$ following similar techniques as those in the NS-$\aa$
\cite{FHTP}.  In particular, we observe that there are two different
power laws for the energy cascade.  For wave numbers $k\ll 1/\aa$,
we obtain the usual $k^{-5/3}$ Kolmogorov power law.  This implies
that the large scale statistics of the flow, in particular for those
eddies of size greater than the length scale $\aa$, are computed
consistent with the Kolmogorov theory for 3D turbulent flows.  On
the other hand,
 for $k \gg 1/\aa$, that is for eddies smaller than the length scale $\aa$, we obtain a steeper power law.  The steeper spectral slope for wave numbers $k \gg 1/\aa$ implies a faster decay of energy in comparison to DNS, which suggests, in terms of numerical simulation, a smaller resolution requirement in computing turbulent flows.  For this reason, we suggest that
 the simplified Bardina model is a good candidate for a subgrid scale model of large eddy simulation of turbulence.  To start, we define some notations:
\begin{eqnarray*}
&&\hskip-.28in
b(u,v,w) = (B(u,v),w),\\
&&\hskip-.28in
\hat{u}_k  = \frac{1}{(2\pi L)^3} \int_{\Om} u(x) e^{-ik\cdot x} \; dx,   \\
&&\hskip-.28in
\hat{v}_k  = \frac{1}{(2\pi L)^3} \int_{\Om} v(x) e^{-ik\cdot x} \; dx,   \\
&&\hskip-.28in
u_k = \sum_{k \leq |j| < 2k} \hat{u}_j e^{ij\cdot x},   \\
&&\hskip-.28in
v_k = \sum_{k \leq |j| < 2k} \hat{v}_j
e^{ij\cdot x},  \\
&&\hskip-.28in
u_k^{<} = \sum_{j < k} u_j,   \qquad v_k^{<} = \sum_{j < k} v_j \\
&&\hskip-.28in u_k^{>} = \sum_{2k \leq j} u_j,  \qquad v_k^{>} =
\sum_{2k \leq j} v_j.
\end{eqnarray*}
There are three flow regimes that we need to consider to analyze the
energy spectra.  These are the flow regimes where energy is
produced, where energy cascades (i.e. inertial range) and, where
energy dissipates and decays exponentially fast (i.e. dissipation
range).  We split the flow into three parts according to the three
length scale ranges.  Assume $k_f < k$, where $k_f$ is the largest
wavenumber involved in the  forcing term.  Thus,
\begin{eqnarray*}
u = u_k^< + u_k + u_k^>\\
v = v_k^< + v_k + v_k^>.
\end{eqnarray*}
The energy balance equation for the simplified Bardina model for an eddy of size $k^{-1}$ is given by
\begin{eqnarray}
&&\hskip-.28in
\frac{1}{2}\; \frac{d}{dt} (v_k, u_k) +\nu (-\Dd v_k, u_k) = T_k -T_{2k},
\label{BAL}
\end{eqnarray}
where,
\begin{equation}
T_k := -b(u_k^<, u_k^<,u_k) + b(u_k + u_k^>, u_k + u_k^>, u_k^<).
\end{equation}\\
We can interpret $T_k$ as representing the net amount of energy per unit time that is transferred into wavenumbers larger than or equal to $k$. Similarly, $T_{2k}$ represents the net amount of energy per unit time that is transferred into wavenumbers larger than or equal to $2k$.  From these definitions $T_k - T_{2k}$ represents the net amount of energy per unit time that is transferred into wavenumbers between $[k,2k)$.\\
Taking an ensemble average (long time average) of (\ref{BAL}) we get:
\begin{equation}\label{TAET1}
\nu \ang{(-\Dd v_k,u_k)} = \ang{T_k} -\ang{T_{2k}}.
\end{equation}
We define the energy of eddy of size $1/k$ as
$$E_\aa(k) = (1 + \aa^2|k|^2)\sum_{|j|=k}|\uhat_j|^2.$$
This definition arose from the fact that we consider
$|u|^2+\aa^2\|u\|^2$ as the ``energy'', since this is the conserved
quantity in the simplified Bardina model equation (see section
\ref{IL}).  Using this definition, we can now rewrite the
time-averaged energy transfer equation (\ref{TAET1}) as
$$
\nu k^3E_\aa(k) \sim \nu \int_k^{2k} k^2E_\aa(k)dk
\sim \ang{T_k}-\ang{T_{2k}}.
$$
Thus as long as $\nu k^3E_\aa(k) << \ang{T_k}$ (that is,
$\ang{T_{2k}}\approx\ang{T_k}$, there is no leakage of energy due to
dissipation), the wavenumber $k$ belongs to the inertial range.
Similar to the other alpha subgrid scale models, it is not known what
is the correct averaged velocity of an eddy of length size $k^{-1}$.
That is, we do not know {\it a priori} in these models the exact eddy
turn over time of an eddy of size $k^{-1}$.  As we will see below, we
have a few candidates for such an averaged velocity.  Namely,
\begin{eqnarray*}
&&\hskip-.28in
 U_k^{0}
 = \ang{ \frac{1}{L^3} \int_{\Om} |v_k|^2
dx}^{1/2} \sim \left(\int_k^{2k}(1+\aa^2k^2)E_\aa(k)\right)^{1/2}\sim
\left( k (1+\aa^2 k^2) E_{\aa} (k) \right)^{1/2},   \\
&&\hskip-.28in
 U_k^{1} = \ang{ \frac{1}{L^3} \int_{\Om} u_k \cdot v_k
dx}^{1/2} \sim \left(\int_k^{2k}E_\aa(k)\right)^{1/2}\sim
\left( k E_{\aa} (k) \right)^{1/2},   \\
&&\hskip-.28in
 U_k^{2} = \ang{ \frac{1}{L^3} \int_{\Om} |u_k|^2
dx}^{1/2} \sim \left(\int_k^{2k}\frac{E_\aa(k)}{(1+\aa^2k^2)}\right)^{1/2}
\sim \left( \frac{k E_{\aa} (k) }{ 1+\aa^2 k^2}
\right)^{1/2}
\end{eqnarray*}
that is,
\begin{equation}\label{n}
U_k^n = \frac{(kE_\aa(k))^{1/2}}{(1+\aa^2k^2)^{(n-1)/2}} \ \ (n = 0,1,2).
\end{equation}
In the inertial range, the Kraichnan energy cascade mechanism states that
the corresponding turn over time of eddies of spatial size $1/k$ with given average velocity as above is about
$$
\tau_k^n := \frac{1}{kU_k^n} =
\frac{(1+\aa^2k^2)^{(n-1)/2}}{k^{3/2}(E_\aa(k))^{1/2}} \ \ (n = 0, 1, 2).
$$
Therefore the energy dissipation rate $\ee$ is
\begin{equation}
\ee \sim \frac{1}{\tau_k^{n}} \int_k^{2k} E_{\aa} (k) dk \sim
\frac{k^{5/2} \left( E_{\aa} (k) \right)^{3/2} }{(1+\aa^2
k^2)^{(n-1)/2}},
\end{equation}
and hence
\[
E_{\aa}(k) \sim \frac{\ee^{2/3} (1+\aa^2k^2)^{(n-1)/3}}{k^{5/3}}.
\]
Note that the kinetic energy spectrum of the variable
$u$ is given by
\begin{eqnarray*}
&&\hskip-.28in
E^u(k)
\equiv
\frac{E_{\aa} (k)}{1+ \aa^2 k^2} \sim \left\{
\begin{array}{ll}   \displaystyle{
\frac{\ee_{\aa}^{2/3}}{k^{5/3}},}
\qquad & \mbox{when  }
k\aa \ll 1\,, \\
\displaystyle{ \frac{\ee_{\aa}^{2/3}}{\aa^{2(4-n)/3}
k^{(13-2n)/3}},} \qquad & \mbox{when  } k\aa \gg 1\,.
\end{array} \right.
\end{eqnarray*}
Therefore, depending on the appropriate average velocity on an eddy of
size $k^{-1}$ for the simplified Bardina model, we would get the corresponding
energy spectra which has a much faster decaying power law $k^{(2n-13)/3}$,
$(n = 0,1,2)$ than the usual Kolmogorov $k^{-5/3}$ power law, in
the subrange $k\aa \gg 1$.  This signifies that the simplified Bardina model,
like the other alpha models, is a good candidate subgrid scale model
of turbulence.
\section{Global Existence and Uniqueness of the Inviscid Simplified Bardina Model}\label{IL}
In this section, we will established the global existence and
uniqueness of the inviscid simplified Bardina model using the
classical Picard iteration method. The inviscid simplified Bardina
modle is equivalent to the functional differential equation
\begin{equation}\label{inv-Bardina}
\begin{split}
\dfrac{dv}{dt} + B(u,u) &= f,\\
v&=u+\aa^2 Au,\\
v(0) = v^{in} &= u^{in}+\aa^2 Au^{in},
\end{split}
\end{equation}
where, for simplicity, we assumed $f$ to be time independent.
\begin{theorem}{\bf (Short time existence and uniqueness)}.  Let $v^{in} \in V', \mbox{ and } f\in V'$.  There exists a short time $T_*(\|v^{in}\|_{V'})$ such that the equation (\ref{inv-Bardina}) has a unique solution $v\in C^1\left([-T_*,T_*],V'\right)$, that is, $u\in C^1\left([-T_*,T_*],V\right)$.
\end{theorem}
\begin{proof}
We will use the classical Picard iteration principle (see, e.g., \cite{Schect}) to prove the short time existence and uniqueness theorem.  Namely, it is enough to show that the vector field $N(v) = f-B(u,u)$ is locally Lipschitz in the Hilbert space $V'$.  From the classical theory of ordinary differential equations we consider the equivalent equation for (\ref{inv-Bardina})
\begin{equation}\label{v-integral}
v(t) = v^{in} -\int_0^t B(u(s),u(s))ds + f\ t.
\end{equation}
Notice that $v\in V'$ implies that $u\in V$ and thus by Lemma
\ref{lem1-buvw} $B(u,u)\in V'$. As a result the equation above makes
sense in the space $V'$.  Let $v_1,v_2\in V'$,  and consequently
$u_1,u_2 \in V$. By (\ref{buvw}) and Poincar\'{e} inequality, we
have
\begin{equation}
\begin{split}
\|N(v_1)-N(v_2)\|_{V'} &= \|B(u_1,u_1)-B(u_2,u_2)\|_{V'}\\
&=\sup_{\{w\in V, \|w\|=1\}} |\ang{B(u_1-u_2,u_2) + B(u_1,u_1-u_2),w}_{V'}|\\
&\leq \dfrac{2c}{\lambda_1^{1/4}}\|u_1-u_2\|\left(\|u_1\|+\|u_2\|\right).
\end{split}
\end{equation}
For any large enough R such that $\|u_1\|,\|u_2\|\leq R$, we have
\begin{equation}\label{lipz}
\|N(v_1)-N(v_2)\|_{V'}\leq \dfrac{4cR}{\lambda_1^{1/4}}\|u_1-u_2\|\leq\dfrac{cR}{\lambda_1}\|v_1-v_2\|_{V'}.
\end{equation}
Here we used the fact that $\|v\|_{V'} \mbox{ is equivalent to } \|u\|$.  Equation (\ref{lipz}) implies that $N(v)$ is locally Lipschitz continuous function in the Hilbert space $V'$.  By the classical theory of ordinary differential equations, the equation (\ref{v-integral}) has a unique fixed point in a small interval $[-T_*,T_*]$ and $v\in C([-T_*,T_*],V')$ (see, e.g., \cite{Schect}). In particular, since $B(u(s),u(s))$ is a continuous function with values in $V'$ and the forcing $f$ assumed to be time independent, equation (\ref{v-integral}) implies that the left hand side $v(t)$ is differentiable and
\begin{equation}\label{inv-Bardina2}
\begin{split}
\dfrac{dv}{dt} &= -B(u,u) + f\\
v(0) &= v^{in}.
\end{split}
\end{equation}
This implies the local-in-time existence and uniqueness of solution $v\in C^1\left([-T_*,T_*],V'\right)$, and hence, $u\in C^1\left([-T_*,T_*],V\right)$,  to the inviscid simplified Bardina model (\ref{inv-Bardina}) or (\ref{inv-Bardina2}).  We will next show that, in fact, we have global existence.  To show global existence to (\ref{inv-Bardina}) or (\ref{inv-Bardina2}) it is enough to show that on the maximal interval of existence, $\|v(t)\|_{V'}$ remains finite.  Let $[0,T_{max})$ be the maximal interval of existence.  If $T_{max} = +\infty$, then there is nothing to prove.  Suppose, for the purpose of contradiction, that
\begin{equation}\label{assumption}
T_{max}<\infty.
\end{equation}
This implies that $\limsup_{t\rightarrow T^{-}_{max}}\|v(t)\|_{V'} = \infty$. By the equivalence of the norms $\|v(t)\|_{V'}\mbox{ and } \|u(t)\|$, we conclude that also
\begin{equation}\label{conclusion}
\limsup_{t\rightarrow T^{-}_{max}}\|u(t)\| = \infty.
\end{equation}
We will derive a contradiction to the conclusion in (\ref{conclusion}).\\\\
Notice that on $[0,T_{max})$, $u\in C([0,T_{max}),V)$, hence we can take the action of (\ref{inv-Bardina}) or (\ref{inv-Bardina2}) on $u(t)$.  We get, by (\ref{buww})
\begin{equation}
\ang{\dfrac{dv}{dt},u}_{V'} = -\ang{B(u,u),u}_{V'} + \ang{f,u}_{V'}= \ang{f,u}_{V'}
\end{equation}
Thus, we have
\begin{equation}\label{above}
\begin{split}
\dfrac{1}{2}\dfrac{d}{dt}\left(|u|^2+\aa^2\|u\|^2\right)&\leq \|f\|_{V'}\|u\|.
\end{split}
\end{equation}
Let $e_0$ be a positive constant which has the same units as $|u|^2$.  From (\ref{above}) we have
\begin{equation}\label{above2}
\begin{split}
\dfrac{1}{2}\dfrac{d}{dt}\left(|u|^2+\aa^2\|u\|^2+e_0\right)&\leq \|f\|_{V'}\|u\|\\
&\leq \dfrac{\|f\|_{V'}}{\alpha}(|u|^2 + \aa^2\|u\|^2+e_0)^{1/2}.
\end{split}
\end{equation}
Denote by $z^2:=|u|^2 + \aa^2\|u\|^2+e_0$.  Then we can rewrite (\ref{above2}) as
\begin{equation}\label{dzdt}
\dfrac{dz}{dt}\leq \dfrac{\|f\|_{V'}}{\alpha}.
\end{equation}
Consequently,
\begin{equation}
z(t)\leq z(0) + \dfrac{\|f\|_{V'}}{\aa}t,
\end{equation}
for all $t< T_{max}$. Therefore, by letting $e_0 \to 0$ we obtain
\begin{equation}\label{lineargrowth}
|u(t)|^2 +\alpha^2 \|u(t)\|^2\leq |u(0)|^2+\aa^2\|u(0)\|+\dfrac{\|f\|_{V'}}{\aa}t,
\end{equation}
and in particular,
\begin{equation}
|u(t)|^2 +\alpha^2 \|u(t)\|^2\leq |u(0)|^2+\aa^2\|u(0)\|+\dfrac{\|f\|_{V'}}{\aa}T_{max}=:K.
\end{equation}
This implies that
\begin{equation}
\limsup_{t\rightarrow T^{-}_{max}}|u(t)|^2+\aa^2\|u(t)\|^2 \leq K.
\end{equation}
 This is a contradiction to the conclusion (\ref{conclusion}).
\end{proof}
To summarize, we have established the proof to the following theorem:
\begin{theorem}\label{inv-thm}
{\bf (Global existence and uniqueness)} Let $f \in V'$ and $v^{in}\in V'$.   Then the system in (\ref{inv-Bardina}) has a unique solution $v\in C^1((-\infty,\infty),V')$ ( or equivalently, $u\in C^1((-\infty,\infty),V)$).
\end{theorem}
We observe that the inviscid Bardina model, (\ref{inv-Bardina}), is equivalent to the following modification of the 3D Euler equations
\begin{equation}
\begin{split}
-\aa^2\Delta\dfrac{\partial u}{\partial t} + \dfrac{\partial u}{\partial t} + (u\cdot \nabla)u + \nabla p &= f\\
\nabla\cdot u &= 0\\
u(x,0)&=u^{in}.
\end{split}
\end{equation}
In particular, it is equal to the Euler equations when $\alpha = 0$.
Therefore, we propose the inviscid simplified Bardina model as
regularization of the 3D Euler equations that could be implemented
in numerical computations of three dimensional inviscid flows. The
analytical study of the regularity of the solutions of the inviscid
simplified Bardina model, and in particular the  limit of its
solutions, as $\alpha \to 0$, to the solutions of the Euler
equations will be reported in a forthcoming paper.

Inspired by the above model, (see also \cite{Kalan, Oskol}), we
propose the following regularization of the 3D Navier-Stokes
equations
\begin{equation}\label{new-mod}
\begin{split}
-\aa^2\Delta\dfrac{\partial u}{\partial t} +
\dfrac{\partial u}{\partial t} -\nu\Delta u+
(u\cdot \nabla)u + \nabla p &= f\\
\nabla\cdot u &= 0\\
u(x,0)&=u^{in}
\end{split}
\end{equation}
subject to either periodic boundary condition or the no-slip
Dirichlet boundary condition $u|_{\partial\Omega} = 0$.  In the
presence of physical boundaries the above regularization
(\ref{new-mod}) of the Navier-Stokes equations is different in
nature from the hyperviscosity regularization of Lions
\cite{Lions2}, or any of the other alpha regularization models,
because it does not require any additional boundary conditions. It
is also simpler than the nonlinear viscosity model of Ladyzhenskaya
\cite{Lad67,Lad85} and Smogarinsky \cite{Sma}.  We will study the
analytical and long-term properties of (\ref{new-mod})  in a
forthcoming paper.

\section*{Acknowledgments}
E.S.T, would like to thank the {\it \'{E}cole Normale Superieure - Paris} for the kind hospitality where this work was completed.  This work was supported in part by the NSF, grants no. DMS-0204794
and DMS-0504619, the BSF grant no. 2004271, the MAOF Fellowship of
the Israeli Council of Higher Education, and the US Civilian
Research and Development Foundation, grant no. RUM1-2654-MO-05.


\begin{thebibliography}{99}
\bibitem{AR75}  R.A. Adams, {\em Sobolev Spaces},
Academic Press, New York, 1975.

\bibitem{Agmon} S. Agmon, {\em Lectures on Elliptic
Boundary Value Problems,} Van Nostrand, New York, (1965).

\bibitem{Bardina} J. Bardina, J. Ferziger, and W. Reynolds, {\em Improved subgrid scale models for large eddy simulation,} American Institute of Aeronatics and Astronautics {\it paper}, {\bf 80} (1980), 80-1357.

\bibitem{BIL} L.C. Berselli, T. Iliescu, W.J. Layton
{\em Mathematics of Large Eddy Simulation of Turbulent Flows,}
Springer, Scientific Computation, New York, (2006).

\bibitem{CI} V.V. Chepyzhov and A.A. Ilyin, {\em On the fractal dimension of invariant set; applications to Navier-Stokes equations,} Discrete Continuous Dynamical Systems, {\bf 10}, (2004), 117-135.

\bibitem{CH98} S. Chen, C. Foias, D.D. Holm, E. Olson, E.S. Titi
and S. Wynne,  {\em The Camassa--Holm equations and turbulence,}
 Phys. D {\bf 133} (1999), no. 1-4, 49--65.


\bibitem{CH99} S. Chen, C. Foias, D.D. Holm, E. Olson, E.S. Titi
and S. Wynne,  {\em A connection between the Camassa-Holm
equations and turbulent flows in channels and pipes,}  Phys.
Fluids {\bf 11} (1999), no. 8, 2343--2353.

\bibitem{CH00} S. Chen, C. Foias, D.D. Holm, E. Olson, E.S. Titi
and S. Wynne,  {\em Camassa-Holm equations as a closure model for
turbulent channel and pipe flow,} Phys. Rev. Lett. {\bf  81} (1998), no.
24, 5338--5341.

\bibitem{CH01} S. Chen, D.D. Holm, L.G. Margolin and R. Zhang,
{\em Direct numerical simulations of the Navier--Stokes alpha
model,}  Phys. D {\bf 133} (1999), no. 1-4, 66--83.

\bibitem{CHOT} A. Cheskidov, D.D. Holm, E. Olson and E.S. Titi,  {\em
On a Leray-$\alpha$ model of turbulence,} Royal Soc. A, Mathematical, Physical and Engineering Sciences, {\bf 461} (2005), 629--649.
\bibitem{CHTi}   C. Cao, D. Holm and E.S. Titi, {\em On the Clark-$\aa$ model of turbulence:  global regularity and long-time dynamics,}  Journal of Turbulence, {\bf 6} (2005), no. 20, 1--11.
\bibitem{CF85} P. Constantin and C. Foias, {\em Global Lyapunov exponents, Kaplan-Yorke formulas and the dimension of the attractors for 2D Navier-Stokes equations,} {\bf 38}, no.1, 1--27,
  Comm. Pure Apl. Math., 1985.
\bibitem{CF88} P. Constantin and C. Foias, {\em Navier-Stokes Equations,}
  The University of Chicago Press, 1988.

\bibitem{CFT} P. Constantin and C. Foias, R. Temam,
{\em Attractors representing turbulent flows}, vii+67, {\bf 53}, no. 314,
  Mem. Amer. Math. Soc., 1985.

\bibitem{FOIAS} C. Foias, {\em What do the Navier--Stokes equations tell us
 about turbulence?} Harmonic analysis and nonlinear differential  equations
(Riverside, CA, 1995), 151--180, Contemp. Math., {\bf 208}, Amer. Math. Soc.,
Providence, RI, 1997.

\bibitem{FHTM} C. Foias, D.D. Holm and E.S. Titi,  {\em
The three dimensional viscous Camassa--Holm equations, and their
relation to the Navier-Stokes equations and turbulence theory,}
J. Dynam. Differential Equations {\bf 14} (2002), 1--35.

\bibitem{FHTP} C. Foias, D.D. Holm and E.S. Titi,  {\em
The Navier--Stokes--alpha model of fluid turbulence. Advances in
nonlinear mathematics and science,} Phys. D {\bf 152/153} (2001),
505--519.

\bibitem{FMRT} C. Foias, O. Manley, R. Rosa and R. Temam, {\em Navier--Stokes Equations and Turbulence,} Cambridge University Press,
Cambridge, 2001.

\bibitem{GH} B. Geurts and D. Holm, {\em Fluctuation effect on 3D-Lagrangian mean and Eulerian mean fluid motion,} Physica D, {\bf 133} (1999), 215--269.

\bibitem{GH2} B. Geurts and D. Holm, {\em Regularization modeling for large eddy simulation,} Physics of Fluids, {\bf 15}, (2003), L13-L16.

\bibitem{Holm-Gibbon} J. Gibbon and D. Holm, {\em Length-scale estimates for the LANS-$\alpha$ equations in terms of the Reynolds number}, Preprint.

\bibitem{JH88} J. Hale, {\em Asymptotic behavior of Dissipative Systems}, Mathematical Surveys and Monographs, {\bf 25},
Amer. Math. Soc. Providence, RI, 1988.


\bibitem{HM}  D. Holm and Marsden, T. Ratiu, {\em Euler-Poincar\'{e} models of ideal fluids with nonlinear dispersion,} Phys. Rev. Lett. {\bf 80}, (1998), 4173--4176

\bibitem{NH} D. Holm and B. Nadiga, {\em Modeling mesoscale turbulence in the barotropic double-gyre circulation,} Journal of Physical Oceanography, {\bf 33} (2003), No.11, 2355-2365.

\bibitem{Hou} T. Y. Hou and R. Li, {\em Dynamic depletion of
vortex stretching and non-blowup of the 3-D incompressible Euler
Equations}, J. Nonlinear Science, (2006) (to appear).

\bibitem{ILT} A. Ilyin, E. Lunasin, E. S. Titi, {\em A modified-Leray-$\alpha$ subgrid scale model of turbulence,} Nonlinearity, {\bf 19}, (2006), 879--897.

\bibitem{Kalan} V.K. Kalantarov, {\em Attractors for some nonlinear
    problems of mathematical physics,(Russian)}
  Zap. Nauchn. Sem. Leningrad. Otdel. Mat. Inst. Steklov (LOMI) {\bf
    152} (1986), Kraev. Zadachi Mat. Fiz. i Smezhnye
  Vopr. Teor. Funktsii18, {\bf 182} 50--54; translation in J. Soviet
  Math. {\bf 40} (1988), no. 5, 619--622.

\bibitem{Kerr}  R. Kerr, {\em Computational Euler history,}
(2006)(submitted). \underline{http://arxiv.org/abs/physics/0607148}.


\bibitem{Lad67} O. A. Ladyzhenskaya, {\em New equations for the description of motion of viscous incompressible fluids and solvability in the large of boundary value problems for them,} Proc. Steklov Insts. Math. {\bf 102}, (1967), 95--118.

\bibitem{Lad85} ------, {\em The Boundary Value Problems of Mathematical Physics,} Applied Mathematical Sciences, vol. {\bf 49}, Springer-Verlag, 1985.

\bibitem{Layton-06}  W. Layton and R. Lewandowski, {\em On a well-posed turbulence model,} Dicrete and Continuous Dyn. Sys. B, {\bf 6}, (2006), 111-128.

\bibitem{LT76}  E. Lieb and W. Thirring, {\em Inequalities
for the moments of the eigenvalues of the Schrödinger
Hamiltonian and their relation to Sobolev inequalities,}
Studies in Mathematical Physics: Essays in Honor of
V. Bargman eds. E. Lieb, B. Simon, and A. S. Wightman,
Princeton University Press, Princeton,
New Jersey, (1976), 226--303.

\bibitem{Lions2} J. L. Lions, {\em Quelques r\'{e}sultats d'existence dans des \'{e}quations aux d\'{e}riv\'{e}es partielles non lin\'{e}aires,} Bull. Soc. Math. France  {\bf 87}, (1959), 245--273.

\bibitem{Lions} J. L. Lions, {\em Quelques M\'{e}thodes de R\'{e}solution Des Probl\'{e}mes aux Limite Non Lin\'{e}aires,} Dunod, Paris, 1969

\bibitem{MM} K. Mohseni, B. Kosovi\'{c}, S. Shkoller and J. Marsden, {\em Numerical simulations of the Lagrangian averaged Navier-Stokes equations for homogeneous isotropic turbulence,} Phys. Fluids, {\bf 15} (2003), No.2, 524--544.

\bibitem{OlTi} E. Olson, E.S. Titi, {\em Viscosity versus vorticity stretching: global well-posedness for a family of Navier-Stokes-alpha-like models}, Math. Anal. Ser. A : Theory Methods, to appear.

\bibitem{Oskol} A.P. Oskolkov, {\em The uniqueness and solvability in
    the large of boundary value problems for the equations of motion
    of aqueous solutions of polymers,(Russian)} Boundary Value Problems of
  Mathematical Physics and Related Questions in the Theory of
  Functions, 7 Zap. Nau\v cn. Sem. Leninggrad. Otdel. Mat. Inst. Steklov. (LOMI) {\bf 38}
  (1973), 98--136.
\bibitem{Pope} S. Pope, {\em Turbulent Flows,} Cambridge University Press, 2001.
\bibitem{JR11} J. Robinson, {\em Infinite-Dimensional Dynamical Systems,} Cambridge University Press, 2001.

\bibitem{MTV} G. Metivier, {\em Valeurs propes d'operateurs definis par la restriction de systemes variationelles a des sous-espaces,} J. Math. Pures Appl. (9) {\bf 57}, 133--56.
\bibitem{Schect} M. Schecter, {\em An Introduction to Nonlinear Analysis,} Cambridge University Press, 2004.
\bibitem{SY11} G. Sell and Y. You, {\em Dynamics of Evolutionary Equations,}
Springer-Verlag, New York, 2002.
\bibitem{Sma} J. Smagarinsky, {\em General circulation experiments
    with the primitive equations. I. The basic experiment,}
  Mon. Weather Rev. (1963), 91--99.

\bibitem{TT84} R. Temam, {\em Navier-Stokes Equations, Theory and Numerical
Analysis,} 3rd revised edition, North-Holland, 2001.

\bibitem{TT88} R. Temam, {\em Infinite-Dimensional Dynamical Systems in
Mechanics and Physics,} Applied Mathematical Sciences, {\bf 68},
Springer-Verlag, New York, 1988.

\bibitem{Townsend} A. Townsend, {\em The Structure of Turbulent Flows,} Cambridge University Press, Cambridge, 1967.

\end{thebibliography}
\end{document}